\documentclass[a4paper]{article}

\usepackage[T1]{fontenc}
\usepackage{ae,aecompl}
\usepackage{fullpage}
\usepackage{natbib}
\usepackage{hyperref}%
\hypersetup{breaklinks,backref,colorlinks,citecolor=blue,linkcolor=blue}

\usepackage{graphicx}	
\usepackage{amsmath}	
\usepackage{amssymb}	
\usepackage{color}
\usepackage{multicol}
\usepackage{multirow} 
\usepackage{bm}        
\usepackage{appendix}
\usepackage[british,english]{babel}

\title
    {Blazars SED in Conical Plasma Flow 
  : a Monte Carlo study}

\date{}

\author
    {Nagendra Kumar$^{1}$\thanks{nagendra.bhu@gmail.com}
  \and  Pankaj Kushwaha$^{2}$ \thanks{pankaj.kushwaha@aries.res.in} 
  \\
    $^{1}$Department of Physics, Indian Institute of Science, Bangalore 560012, India\\
    $^{2}$Aryabhatta Research Institute of Observational Science (ARIES), Nainital 263002, India}
    
\let\OLDthebibliography\thebibliography
\renewcommand\thebibliography[1]{
  \OLDthebibliography{#1}
  \setlength{\parskip}{0.1pt} 
  \setlength{\itemsep}{0pt plus 0.3ex}
}

\begin{document}


\maketitle




\begin{abstract}
Blazars host the most powerful persistent relativistic conical jet-- a highly
collimated anisotropic flow of material/plasma. Motivated by this, we explore the
blazar's broadband spectral energy distribution (SED) in an anisotropic flow of
plasma which emits via synchrotron and inverse Compton (IC) mechanism. The
flow is conical with two velocity components: a highly relativistic flow component
along the jet axis and a random perpendicular component with average random Lorentz
factor $\langle \gamma^{ran} \rangle$ {\(<<\)} than the average component along
the jet axis $\langle \gamma \rangle$. Assuming a broken power-law electron population, we calculated the broadband SED using synchrotron and IC processes assuming a
cylindrical (of radius R and length L) emission region. For the IC process, we used
Monte Carlo approach. We found that such anisotropic flow can reproduce blazars
broadband emission as well as general short and high amplitude variability, indicating
that spectral and temporal variability are not sufficient to distinguish among
existing models. We demonstrate this by reproducing SEDs of FSRQ 3C 454.3 and three
BL Lacs objects OJ 287, S5 0716+714, PKS 2155-304. Our formalism and set-up also
allow us to investigate the effect of the geometry and dimension of emission region
on observed broadband spectra. We found that the SEDs of low synchrotron peak (LSP)
blazar can be explained by considering only SSC (synchrotron self-Compton) if R/L ($<$ 0.01), broadly mimicking a spine-sheath geometry. In general, the degeneracy between
non-thermal particle number density and length of the emission region (L) allow us
to reproduce any variability in terms of particle number density.\\

\end{abstract}


\section{Introduction}
Blazar, comprising of BL Lacartae objects (BLLs) and flat spectrum radio quasars
(FSRQs), is a subclass of active galactic nuclei with a highly collimated
relativistic jet of plasma aligned at close angles to our line of sight. They are
characterized by a high and rapidly variable continuum, exhibiting a broad double-humped spectral
energy distribution (SED, $\nu$ vs $\nu F_\nu$) extending from radio to GeV/TeV
$\gamma$-ray energies. The low-energy hump extends from radio to ultraviolet (UV)/X-ray
with a maximum in between near-infrared (NIR) to UV/X-rays while the high-energy
hump spans X-rays to GeV/TeV $\gamma$-rays with a maximum at sub-GeV to GeV
energies. The radio and optical emission, which constitute the low-energy hump
show high and variable polarization, and thus, is widely regarded as synchrotron
emission from the relativistic non-thermal electrons in the jet. The characteristic
double-humped SED and an apparent positive correlation between the maximum of the
two humps have led to an additional classification scheme based solely on this
\citep{Fossati-etal1998}. Thus, based on the location ($\nu_s$) of the maximum of
the low-energy hump, blazars have been termed low-synchrotron peaked (LSP,
$\nu_s \leq 10^{14}$ Hz), intermediate-synchrotron peaked (ISP, $10^{14} <
\nu_s < 10^{15}$ Hz), and high-synchrotron peaked (HSP, $\nu_s > 10^{16}$ Hz).
However, so far only BLLs have been found to conclusively show the three spectral
subclassed, referred respectively as LBL, IBL, and HBL while FSRQs are exclusively
LSP.

For the high-energy hump, two competing scenarios, broadly refereed to as leptonic
and hadronic,
depending on the primary particles responsible for the high energy emission have been
proposed in the literature. In the leptonic scenario, entire blazar emission is due
to electrons (or positrons) and it is one of the widely considered scenario due to
presence of the synchrotron emitting relativistic electrons and a wide range of soft
photon fields from different AGN constituents. It attributes the X-rays and $\gamma$-rays
to the up-scattering of the local soft-photon fields by electrons through inverse Compton
(IC) mechanism \citep[e.g.][]{Ghisellini-2009,Kushwaha-etal2013,Kushwaha-etal2018a}.
The widely considered soft-photon fields include synchrotron photons
in the jet and photons external to the jet e.g. from AGN constituents like accretion
disk (AD), broad line region (BLR), IR torus (IR), and the omnipresent cosmic microwave
background (CMB) photons. Accordingly, their respective IC spectra are generally
refereed in literature as synchrotron self-Compton (SSC), EC-AD, EC-BLR, EC-IR, and
EC-CMBR where EC stands for external Comptonization -- IC of photon fields external
to the jet. Depending on the location of emission region, one or many of the 
soft photon fields may contribute significantly to the observed emission. The alternative
scenario based on hadronic interactions attributes the
high energy hump to proton-synchrotron and/or cascades initiated as a result
of hadron-hadron and hadron-photon interactions \citep[e.g.][and references therein]
{Bottcher-etal2013}.

Based on these two scenarios, a variety of models, time-dependent and time-independent
have been proposed in the literature for the double-humped SEDs. The
time-dependent model additionally provides insights into the likely jet physical
conditions and processes causing the observed changes by attempting to reproduce
additional observational quantities like light curves \citep[e.g.][]{,Kushwaha-etal2014}, polarization evolution \citep[e.g.][]{,Chandra-etal2015} etc. or the general
observational trends \citep[e.g.][]{Marscher2014}. The time independent
model, on the other hand, only concerned with reproducing the observed SED, and
can be broadly considered as a snapshot moment of the time dependent models.

A huge range of
rarely repeating observational behaviors exhibited by these
sources combined with enormous scale of separation between the extremes
suggest a very diverse and extreme jet physical conditions. On one hand,
magnetic field is believed to play an important and dominant role in the formation
and collimation of relativistic jets close to the central engine, broadband SEDs
modeling suggests a kinetically dominated jet. Despite the importance of magnetic
field in jets, its configuration is still uncertain, however the polarization
measurement for low energy hump emission (or synchrotron radiation) suggest, in
many cases, a helical magnetic field configuration \cite[e.g.,][]{Lyutikov-etal2005, Gabuzda2017}. Further, though we have a general understanding and idea of relevant 
physical conditions and radiation processes in the jet, we still lack a comprehensive
understanding. All the models proposed in the literature for emission or observed
behaviors are primarily specific, with limited applicability-- capable of explaining
a few of the observed features/trends or explain the general features e.g. stochastic
variability \citep[e.g.][]{Marscher2014} etc.
Specifically, apart from the beamed emission it is expected that the emission model 
must explain the observed features like variability time scale and at the same
time, should be optically thin to pair production \cite[see,][and references therein]{Maraschi-etal1992, Dondi-Ghisellini1995} given the fact that the 
shortest variability time scale provides a maximum bound on the size of the emission
region, which is usually order of hours to days.

From the observer
point of view, as the jet appears in radio images, its a conical collimated flow
i.e. an anisotropic flow. The strong kinetic dominance inferred in a few cases
\citep{Algaba-etal2019} suggests an anisotropic flow locally. %
 Whether such a local anisotropic flow can emit powerfully can be determined
by its capability to produce the emission with observed features  within the
known observational constraints. 
Also how it accelerates anisotropically, is still an open issue. In literature, attempts have been made to
understand an anisotropic particle acceleration \cite[e.g.,][]{Armstrong-etal2012,
Yang-Zhang2018}. In this work, we focus solely on the emission aspect with
assuring that within the emission region the flow is stable in assumed magnetic field configuration$-$ wiggler plus axial magnetic field. 
In helical undulator (or free-electron laser) principle, the anisotropically moving relativistic electron emits THz radiation (even X-ray emission, e.g., \citet{Pellegrini-etal2016}) in the presence of wiggler plus axial magnetic field (verified in laboratory).  
Motivated from the helical undulator principle we assume that 
the conically moving electron will emit synchrotron radiation in presence of a wiggler
(acting in plane perpendicular to the jet axis) plus axial magnetic field.
Though the formulation of helical undulator is non-linear it assures (proven
experimentally) that when one of electron velocity component is
along the wiggler field direction and another one is along the axial field
direction then electron moves on stable orbit along the wiggler field path
without changing the magnitude of both  the velocity components \cite[e.g.,][and references therein]{Ginzburg-Peskov2013, Balal-etal2015}.
However, a polarization measurement for synchrotron radiation in
wiggler
+ axial magnetic field  is needed to establish the viability of this field
configuration in the jet environment. We intended to study this in future.
In this work, we explore this scenario by considering a local conical flow
and model the SED assuming leptonic emission processes. We show that the
synchrotron emission is beamed (but not due to a bulk Doppler boosting) and
IC process dominates over the pair production process. 
In the next
section, we present the details of model and geometry of the emission region.
\S\ref{sec:SEDcalc} presents the SED modeling approach and \S\ref{sec:SEDcomp} discusses
the comparison of observed SED of blazars followed by our summary and conclusions
in \S\ref{sec:sumCon}. 

\section{Emission mechanism of the conical jet model 
}

\begin{figure} 
\centering
\begin{tabular}{lr} 
  \includegraphics[width=0.36\textwidth]{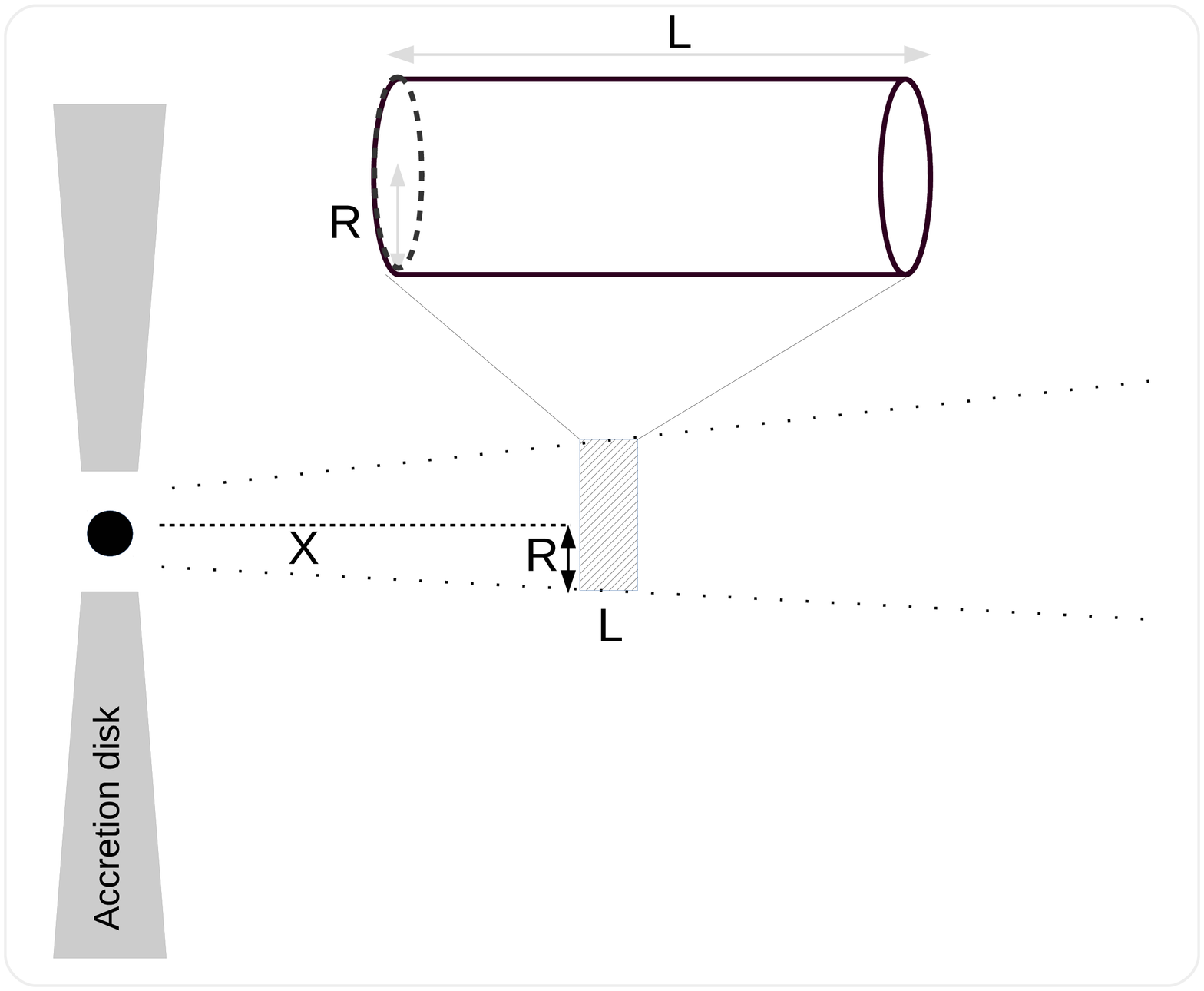} &\hspace{.20cm}
  \includegraphics[width=0.36\textwidth]{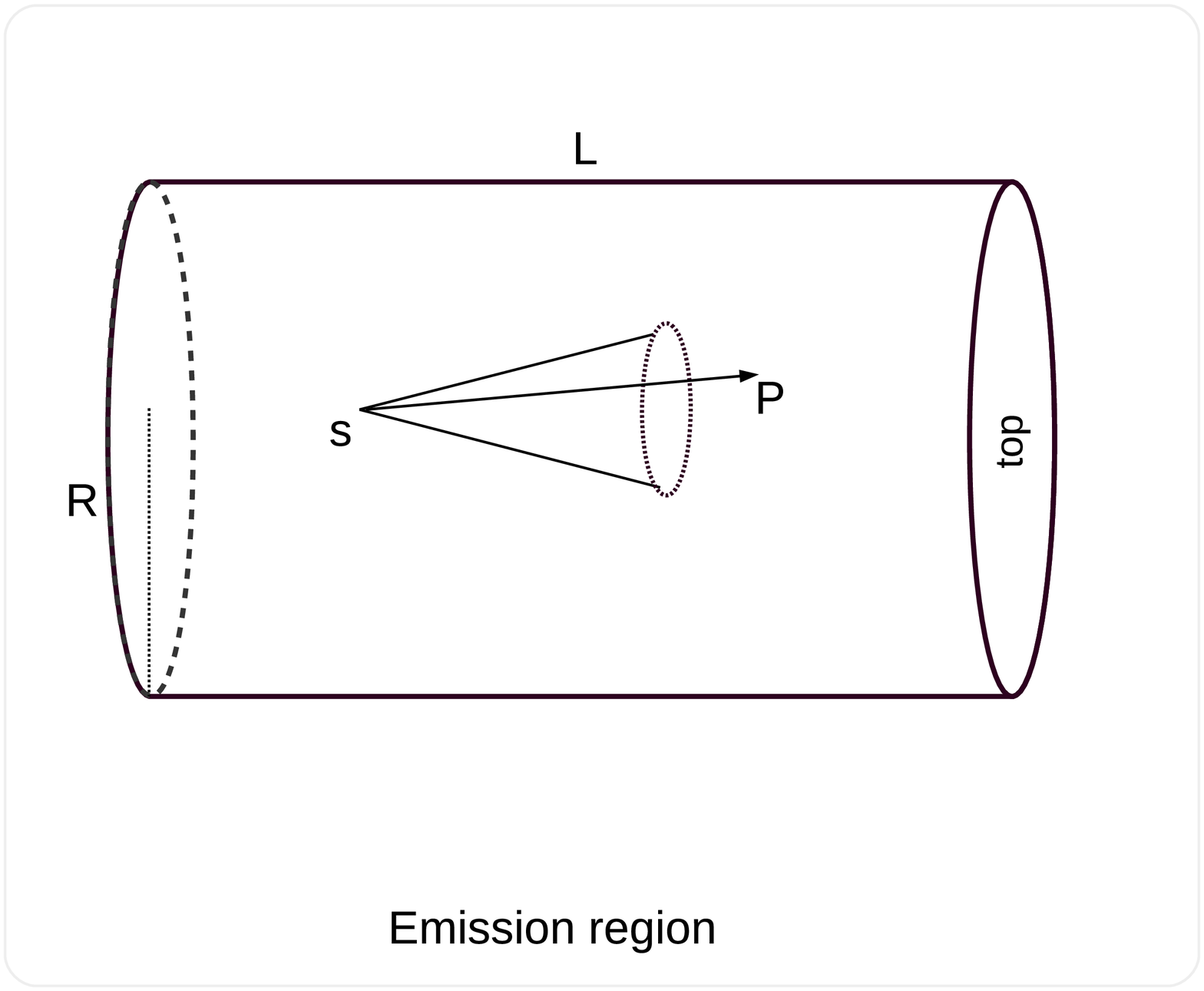} \\
\end{tabular} 
\caption{{\it Left:} A cartoon diagram of a conical jet with a semi aperture angle
of $\theta_j$ and the embedded emission region marked by a shaded area of length
L. The emission region is at X distance from the central engine. Since L $\ll$ X, the shape
of emission region can be approximated to a cylindrical (its zoomed version has
been shown, where R is the radius and L is the length of the cylinder). {\tt Right:} The motion
of jet-matter inside the cylindrical emission region. At any point s inside the emission
region, the flow lies inside the cone of angle $\theta_j$. One such
directions on the surface of cone is shown by the arrow {\tt sp} (see Kumar
2018). The face marked {\tt top} is the face of emission region towards the observer.  
} 
\label{fig:blz-ma}
\end{figure}

Motivated by the observed jet morphology in radio observations, we assumed a conical
jet of opening angle or semi-aperture angle $\theta_j$, as shown in figure
\ref{fig:blz-ma}, where the emission region (responsible for the observed emission) is embedded in the flow. The emission region is located X distance from the central engine (shaded region in figure \ref{fig:blz-ma}) and usually lie beyond the acceleration region. Within  it electrons have attained the relativistic conical flow now steadily. Our primary goal is to generate the SEDs of blazar in leptonic model having a steady conically flowing relativistic electron inside the emission region.
The maximum size of emission region L is usually estimated from the flaring phase of
blazars and  is order of $\gamma$-ray variability time scale ($\sim$ hours to days).
As per current understanding from spectral and temporal studies, blazar's emission
region is widely agreed to be mainly at sub-parsec or parsec scales i.e. $\rm
X\gtrsim 0.1~pc$ \cite[e.g.][]{ Marscher2006, Ghisellini-2009, Agudo-etal2013, Ghisellini-etal2014}.  Thus, for a compact
emission region (L $\ll$ X), one can approximate the conical emission region ($\rm R_{top}$ > R) 
to a cylindrical ($ \rm R_{top}$ = R, here R and $R_{top}$ are the radius of emission region of the base and top face respectively, and $\rm \frac{R_{top}}{R} = 1 + \frac{L}{X}$\big). 
Without any loss of generality, in our calculation we have considered a cylindrical emission region of radius R and length L (as shown in figure \ref{fig:blz-ma}). 

Since the flow is conical, it has two velocity components: one along the jet axis
which is the dominant one and the other perpendicular to it. The corresponding Lorentz factors are $\gamma$ and $\gamma^{ran}$ respectively (i.e., the
particle is moving mainly along the jet axis with Lorentz factor $\gamma^R$ which corresponds to its composite velocity  v$_R$ of both components v and
 v$_{ran}$, as  v$_R$ = $\sqrt{{\rm v^2 + v}_{ran}^2 - {\rm v^2v}_{ran}^2/c^2}$ or $\gamma^R$ = $\gamma \ \gamma^{ran}$ \big). 
The direction of dominant velocity component is defined same to that in \citet[][]{Kumar2018} \cite[see also,][]{Kumar2017}, and one of its directions is
shown by sP vector in the right panel of figure \ref{fig:blz-ma}.  The other
component lies in the plane perpendicular to the vector sP and points randomly
in this plane (not shown here for clarity). Our main focus here is to explore
the spectral and temporal behavior in an anisotropic flow \cite[e.g.,][]{Yang-Zhang2018},
dominated by the velocity component along the jet axis and all quantities refer
to the measurement in the observer's frame.

\subsection{Synchrotron emission}

The anisotropically moving relativistic electrons will emit the synchrotron emission
in presence of magnetic field and also up-scatter the local soft photon fields via IC.
In many observations, the jet polarization indicates that the flow carries helical magnetic fields \cite[e.g.,][]{Lyutikov-etal2005, Gabuzda2017}.
As discussed in the introduction,
the helical periodic magnetic field (wiggler, $B_w$(\textbf{i} $\cos(k_w z)$+\textbf{j} $\sin(k_w z)$), here $B_w$ is the amplitude of wiggler field and $2\pi/k_w$ is the wiggler period)  along with axial magnetic field (\textbf{k} $B_o$, here {\bf i, j, k} are the cartesian axis where {\bf k}-direction along the jet-axis) is a promising field configuration to generate the synchrotron emission by anisotropically moving electrons, and has been extensively studied in
the case of a helical undulator \cite[e.g.,][]{Balal-etal2015, Ginzburg-Peskov2013}.

Similar to the synchrotron emission mechanism in helical undulator, we consider that the random velocity component (v$_{ran}$) of electrons is along the wiggler field direction. In that situation electron will move on stable helical trajectory (which is along the wiggler field) with constant axial speed (i.e., dominant velocity component) and having constant of motion,  $\gamma^R m_e c$ = constant.
 In the appendix 
we have revisited the single electron trajectory in a  wiggler + axial field.
The synchrotron radiation is emitted in forward direction with frequency $(\gamma^R)^2 \nu_b$ (where $\nu_b = \frac{eB_o}{2\pi m_e c}$ is a Larmor frequency, $e$ is the electronic charge, $m_e$ is the mass of electron, $ c $ is the speed of light).
 The resulting synchrotron radiation will be beamed in the forward direction
  along the jet axis. 
The efficiency of emission depends on the ratio of cyclotron frequency ($\nu_b \gamma^R$) to the bounce frequency ($\Omega_b$ = v$k_w$)
\cite[e.g.,][]{Balal-etal2015, Ginzburg-Peskov2013}. 
 Here, we would stress that for SSC spectrum calculation we consider observed synchrotron emission as a seed photons and $\gamma^R$ is determined from the observed SED shape. In another way, here, we surmise that wiggler + axial magnetic field configuration in jet environment is efficient to generate the synchrotron emission by anisotropically moving relativistic electron.

As per the observed blazars SED shape (low-energy hump), we assumed a broken powerlaw
distribution for a synchrotron emitting electrons, defined by

\begin{equation*}\label{eq:partdist}  
  N'(\gamma^R)d \gamma'=\left\{
  \begin{array}{l r}
    K (\gamma^R)^{-p} d \gamma' & \quad ;\gamma'_{min}< \gamma^R \leq \gamma'_b \\
    K \left.\gamma'_b\right.^{q-p} (\gamma^R)^{-q} d \gamma' & \quad ;\gamma'_b \leq \gamma^R < \gamma'_{max}\\
  \end{array}\right.
\end{equation*}

\noindent where, the electron Lorentz factor $\gamma^R$ is in the range of $\gamma'_{min}$
to $\gamma'_{max}$ and $\gamma'_b$ corresponds to the break in the distribution. p and
q are the powerlaw indices of the electrons, before and after the break respectively.
K is the normalization factor which is related to the total electron density inside
the $\gamma$-ray emitting region as

\begin{equation}\label{k1-value}
  n_e = \frac{K}{p-1} (\left.\gamma'_{min}\right.^{1-p} - \left.\gamma'_{b}\right.^{1-p})
\end{equation}  

With $q > p$ and p $>$ 1, as is the case with blazars SEDs, if $\gamma'_{b} \gg \gamma'_{min}$
then $n_e$ $\left(\approx\frac{K}{p-1} \left.\gamma'_{min}\right.^{1-p}\right)$ will depend on
$\gamma'_{min}$ and $p$. In our model (i.e., in the observer's frame) the peak frequency of synchrotron spectrum $\nu^{syn}_p$ is given as 
\begin{equation}\label{syn-peak}
  \nu^{syn}_p = \frac{\left.\gamma'_b\right.^2}{1+z} \nu_B
  \end{equation}
with $z$ being the redshift of the source. Similarly the synchrotron frequency corresponding to $\gamma'_{min}$ and $\gamma'_{max}$ is defined as $\nu^{syn}_{min} = \frac{\left.\gamma'_{min}\right.^2}{1+z} \nu_B$, $\nu^{syn}_{max} = \frac{\left.\gamma'_{max}\right.^2}{1+z} \nu_B$ respectively.

Further, we assumed that the magnitude of the random velocity component of electron is constant for simplicity. Since $\gamma^R$ = $\gamma \ \gamma^{ran}$, the forward velocity
component of
electron will follow also the broken powerlaw distribution 
as given below
\begin{equation}\label{eq:partdist_f}  
  N(\gamma)d \gamma=\left\{
  \begin{array}{l l}
    K \gamma^{-p} d \gamma & \quad ;\gamma_{min}< \gamma \leq \gamma_b \\
    K \gamma_b^{q-p} \gamma^{-q} d \gamma & \quad ;\gamma_b \leq \gamma < \gamma_{max}\\
  \end{array}\right.
\end{equation}

\noindent  where, the Lorentz factor of electron's forward conical velocity component $\gamma$ is in the range of $\gamma_{min}$
to $\gamma_{max}$. $\gamma_b$ corresponds to the break in the distribution and these are related to the electron Lorentz factor as $\gamma'_{min}$ = $\gamma_{min} \ \gamma^{ran}$; $\gamma'_{b}$ = $\gamma_{b} \ \gamma^{ran}$; $\gamma'_{max}$ = $\gamma_{max} \ \gamma^{ran}$. Similarly, the synchrotron frequency is expressed as $\nu^{syn}_p = \frac{\gamma_b^2(\gamma^{ran})^2}{1+z} \nu_B$; $\nu^{syn}_{min} = \frac{\gamma_{min}^2(\gamma^{ran})^2}{1+z} \nu_B$. 
The average Lorentz factor for dominant velocity component of electron  
$\langle \gamma \rangle$ is given by $\langle \gamma \rangle$  = $\frac{\int \gamma N(\gamma) d\gamma}{\int N(\gamma) d\gamma}$.
Since, we are interested  in exploring blazar's SED from an anisotropic motion
of electrons, to ensure the anisotropic motion of electron we fix
$\gamma^{ran}$ to $\gamma_{min} /2$.
Further, since the electron Lorentz factor $\gamma^R$, also $\gamma$, follows the broken powerlaw distribution, for generality (without affecting the results, see figure \ref{blz-gen}b of subsection \ref{subsec:321}) we assume that  
the random velocity component follows a uniform distribution in range [$\gamma^{ran}_{min}, \gamma^{ran}_{max}$] with $\gamma^{ran}_{max}$ = $\gamma_{min}$ and $\gamma^{ran}_{min} \ll \gamma_{min}$, thus the mean  $\langle \gamma^{ran}\rangle$ $\left(=\frac{\gamma^{ran}_{min}+\gamma^{ran}_{max}}{2}\right)$ $\approx$ $\gamma_{min}/2$.

For a well observed SED from mm to $\gamma$-ray energies, the low-energy hump being
synchrotron emission provides a direct relation to infer the particle spectral
indices if NIR-optical part is not heavily contaminated by the thermal components
(e.g. BLR and IR torus for LSP sources). However, currently we have better coverage
and observational cadence in NIR-optical and often in X-rays. Since the NIR to
X-ray emission, for all the three blazars spectral classes, covers emission before
and after spectral break in the leptonic emission scenario, the particle spectral
indices can be inferred fairly well from spectrum in these parts \citep[e.g. ][]{Kushwaha-etal2013}.

\subsection{IC process and Monte Carlo Method for IC spectrum}
The presence of a diverse range of soft photon fields from different AGN constituents
and the relativistic synchrotron emitting electrons provide a natural environment
for high energy emission via IC scattering. We assumed this scenario for the
broadband emission, i.e. entire emission due to synchrotron and IC.
Since, the conically moving relativistic electrons emit the synchrotron radiation in forward direction (in presence of wiggler + axial magnetic field), we consider  
 a conical direction with same semi aperture angle  $\theta_j$ (in similar way as described in figure \ref{fig:blz-ma}) for a synchrotron photon direction. That is, for a given scattering site, the direction of 
synchrotron photon is any one of directions on the surface of cone at that
site. 
  The optical depth ($\tau$) for IC (in cylindrical emission region) is defined along the jet-axis, i.e., $\tau$ = L $n_e \sigma_T$ = L/$\lambda$, here $\sigma_T$ is the Thomson cross section, $\lambda$ is the mean free path of photons. We have computed the mean free path of photons considering only forward conical velocity component of electron  for SSC. 

For  EC, we assumed an isotropic
external photon filed approximated by a blackbody. Using a more elaborate
geometry of external photon filed doesn't change the outcome significantly
\citep[e.g.][]{Bottcher-etal2013}. The mean free path of photons has been computed  considering the composite velocity of electron for EC.
We are computing the time-averaged IC spectrum over the time L/$c$, we have populated the scattering event homogeneously inside the emission region. 
However, 
  we compute the time-averaged IC spectrum 
  for both flaring and quiescent phase spectrum. 
  Physically,
after the time L/$c$, electrons would be cooled by both synchrotron and IC process
and one can expect the evolution of the spectra from flaring phase to quiescent. In
present study, to avoid the complexity, the evolution of spectra has not been
studied, which we are intended to study in future.

In the present scenario, in 
IC process the photons get up-scattered by both the velocity
components (\citealp[first formulated by][]{Blandford-Payne1981}; \citealp[for
high energies X-ray emission mechanism in X-ray binaries,][]{Titarchuk-etal1997, Kumar2017}) of electron. 
To estimate the
IC spectrum from different photon fields, we followed the Monte Carlo, MC, approach developed
by \citet[and references therein]{Kumar2018, Kumar2017}, with
appropriate modifications for the relativistic scattering regime.
Further, in case of IC calculation (only for computational simplicity), we have allowed the random component to be in any plane rather than restricting in plane perpendicular to the forward velocity vector.
Since, if one computes the IC spectrum considering the random velocity component either strictly in the perpendicular plane of the axial velocity or any plane, then in both cases the IC spectrum will be similar.   
In this work, we have computed the spectrum for fixed opening angle of conical flow, i.e., jet opening angle.
  
      For the assumed geometry, flow and seed photon fields, the rear-end scattering will dominate for SSC process while the isotopic scattering in EC process.
In rear-end IC scattering, photon gains lesser energy compared to the head-on scattering. Thus, it is expected that for modeling the same SED of blazar the required Lorentz factor for SSC would be much larger in comparison to
the respective EC dominated values (see, \S \ref{subsec:ECdom}, \S \ref{subsec:onlySSCFS}). 

Electron direction can change after a scattering in IC process, which may alter the synchrotron emission  or conical flow's direction. 
In IC process, for highly relativistic electron ($\gamma \gg$ 1) and low-energetic unscattered photon ($E_p \ll m_e c^2$, $E_p$ is the photon energy), which is our considered case, the scattered
photon is always  in the unscattered electron's direction. 
In addition, the scattered elctron and photon are in same 
direction (\citealp[][]{Pozdnyakov-etal1983}, see also \citealp{Kumar-Misra2016}).
Hence, the electron direction does not get alter after IC process in our framework, i.e., it will maintain the conical flow, and the synchrotron emission is always beamed in forward direction. 

The spectral peak frequency of SSC in the assumed
scenario for single scattering is given by $\nu_p^{ssc} \propto \langle
\gamma^2 \rangle \langle \gamma^{ran} \rangle^2 \langle \nu_p^{syn} \rangle$, where
$\langle \nu_p^{syn} \rangle$ = $\langle \gamma^2 \rangle\langle \gamma^{ran} \rangle^2 \nu_B$ is an average seed
synchrotron photon frequency for SSC;
and $\langle \gamma^2 \rangle$ = $\frac{\int \gamma^2 N(\gamma)
  d\gamma}{\int N(\gamma) d\gamma}$.
Here, we consider a proportionality relation for $\nu_p^{ssc}$ \big(usually, $\nu_p^{ssc} \approx \left(1-\frac{\rm v}{c}\right)^2 \langle \gamma^2 \rangle \langle \gamma^{ran} \rangle^2 \langle \nu_p^{syn} \rangle $ for rear-end scattering \big), mainly because of that $\nu_p^{ssc}$ also depends on $\theta_j$ of conically moving relativistic electron (as, \citealp{Kumar2018} has shown that IC spectra for rear-end scattering becomes harder by increasing $\theta_j$ for a given electron energies).
 We have checked the proportionality relation for $\nu^{SSC}_p$ in two ways in Monte Carlo method. In first case we change $\langle \gamma^{ran} \rangle$ in such a way that $\langle \gamma^2 \rangle$ is unchanged (see figure \ref{blz-gen}c of subsection \ref{subsec:321}) and in second we change $\gamma_{min}$ (see figure \ref{blz-cylinder}c of subsection \ref{subsec:322}). 
In case of EC process, the IC scattering is an isotropic so the scattered frequency would be independent of $\theta_j$. The peak frequency of EC $\ \nu_p^{EC}$ for single scattering is simply determined by the relation $\nu_p^{EC} \approx \langle \gamma^2 \rangle \langle \gamma^{ran} \rangle^2  \nu_p^{bb}$, here $\nu_p^{bb}$ is the peak frequency of the blackbody emission at temperature $T_{EC}$ (see  subsection \ref{subsec:ECdom} and table \ref{tab-fl} for MC results).

\subsection{Optically thin IC spectrum calculation}

In the IC scattering, the mean free path of photons inside the electron
scattering medium increase with electrons speed. Especially, for $\gamma$ $>$ 10, $\lambda$
increases rapidly, e.g., $\lambda$ is $\sim$ 6.0$\times 10^{17}$, 5.8$\times 10^{19}$,
5.7$\times 10^{21}$ cm for $\gamma$ = 10, 100, 1000 respectively for $10^{15}$ Hz photon at $n_e$ = 
10$^{8}$ cm$^{-3}$ \cite[e.g.,][]{Kumar2018}. In general, the mean free path of the
particle in medium is inversely proportional to the medium density and total differential
cross-section. Further, for a given electron population there exists a minimum
average scattering number ($\langle N_{sc}\rangle$) after which the photons energy
spectrum does not change with increasing $\langle N_{sc}\rangle$. This resulting
limiting photons spectra are referred as Wien's peak spectra. Similar to the case
of variation of mean free path, the Wien peak spectrum in ultra relativistic limit
occurred for a low value of minimum $\langle N_{sc} \rangle$ (e.g., the
minimum $\langle N_{sc} \rangle$ $\sim$10, $\sim$6 and $\sim$2 for $\gamma$ = 20, 100
and 1000 respectively), while in non-relativistic limits Wien peak is observed for a large value of
$\langle N_{sc} \rangle$ \cite[e.g.,][]{Kumar-Misra2016}. In this work we will
compute/compare the SED almost single scattering limit, i.e., very far from the Wien's peak spectrum limit, or in optically thin regime. 
The MC treatment already takes all the special relativistic effect. We only
need to apply the redshift to the emitted spectrum to transform it to the observed
one.

\subsection{The pair production}
Pair production is an another competing process which may affect observed SEDs as
the $\gamma$-ray produced by Compton process will get disappeared by producing $e^+e^-$
pairs if the threshold energy value for the two photons ($E_{p1}$, $E_{p2}$) becomes
$\geq2m_e c^2$ i.e. $E_{p1} + E_{p2} \geq 2 m_e c^2$ where $m_e$ is the mass of
an electron/positron, and $E_{p1} = h\nu_{p1}$ with $h$
being the Planck constant. The cross section for pair production is given by
\cite[e.g.,][]{Aharonian-etal2008, Pozdnyakov-etal1983}  

\begin{footnotesize}
\[\sigma_{\gamma \gamma} = \frac{3 \sigma_T}{8 s}\left[\left(2+\frac{2}{s}-\frac{1}{s^2}\right) \ln (\sqrt{s}+\sqrt{s-1}) - \left(1+\frac{1}{s}\right)\sqrt{1-\frac{1}{s}}\right] \]
\end{footnotesize}

\noindent Here s=$\dfrac{E_{p1}E_{p2}}{m_e^2c^4}$ $\approx$ $\dfrac{\nu_{p1} \nu_{p2}}{1.52 \times 10^{40} Hz^2}$
and the above mentioned threshold corresponds to $s=1$. Thus, $\sigma_{\gamma \gamma}$
= 0 for all s $\leq$ 1 (threshold) and at s = 2 the cross section for pair production has
maximum $\sigma_{\gamma \gamma}$ = 0.25$\sigma_T$. From threshold s = 1, we can see
that photons of frequency less than $10^{12}$ Hz needs a photon of energy $>10^{28}$ Hz
($\sim 41$ TeV) to produce $e^+e^-$ pair, but observed spectra of a majority of
blazars normally extend up to a few hundreds of GeVs. The other crucial factor
is the number density of the soft photons which, in general, follow a monotonic
declining trend with some powerlaw form or steeper from NIR to GeV/TeV energies. For example, when the electron energy distribution has single powerlaw distribution with index p, the synchrotron
photon density is proportional to $\nu^{-(p-1)/2}$ \cite[e.g.,][]{Shah-etal2017}, so for p=2 the synchrotron photon
density at 10$^{14}$ Hz (or, $\geq$ 10$^{14}$ Hz) is almost 10 times smaller than that at 10$^{12}$ Hz (or, $\geq$ 10$^{12}$ Hz). As a
result, photons of frequency 10$^{26}$ Hz will see 10 times smaller photon density
for pair production in comparison to the 10$^{28}$ Hz photons.
Other way around, the probability for pair production by 10$^{26}$ Hz photons
via photon-photon scattering is almost 10 
times lesser when it is done by 10$^{28}$Hz. Further, the cross section
for pair production ($\sigma_{\gamma \gamma}$) is smaller than IC cross section.
 Thus, in the optically thin regime i.e. in the single scattering limit,
the pair production will be negligible.

\section{General Results and SED Calculation}\label{sec:SEDcalc}
In the leptonic emission scenario, the location of emission region plays a
crucial role in deciding the role of external photon fields as seed photons
for IC. In general, it is inferred from the modeling of the SEDs \citep[e.g.][]{Ghisellini-2009,Kushwaha-etal2013}. Rather than deriving the temperature
from SED fitting, we tested two temperature -- 500 K and 50k K, representative
of IR torus and BLR field. 

As it turned out, for the assumed cylindrical geometry of the emission region, its length (L) vis-a-vis mean free path ($\lambda$) has strong effect on the
direction of emitted synchrotron photons and hence, SSC as well. The next subsection
(\S\ref{subsed:synIC}) explore this aspect. Other uncertainties in the model are
effects of use of limited energy range of synchrotron emission for calculating SSC spectrum,
 arbitrariness of the subdominant random velocity components on
the model generated IC spectrum, and the effect of dimension of emission region.
We investigate these in the subsection \S\ref{subsec:32}. It should be noted that
in both these subsections, we do not focus on constraints from blazar observations
but only on the effects of these uncertainty on the model generated spectrum. After
these assessment, we consider modeling blazar SEDs using observational
constraints in \S\ref{subsec:blazarSEDs}.

\subsection{Estimation of observed synchrotron emission}\label{subsed:synIC}

\begin{figure} 
\centering
\begin{tabular}{lr}\hspace{-0.8cm}
  \includegraphics[width=0.46\textwidth]{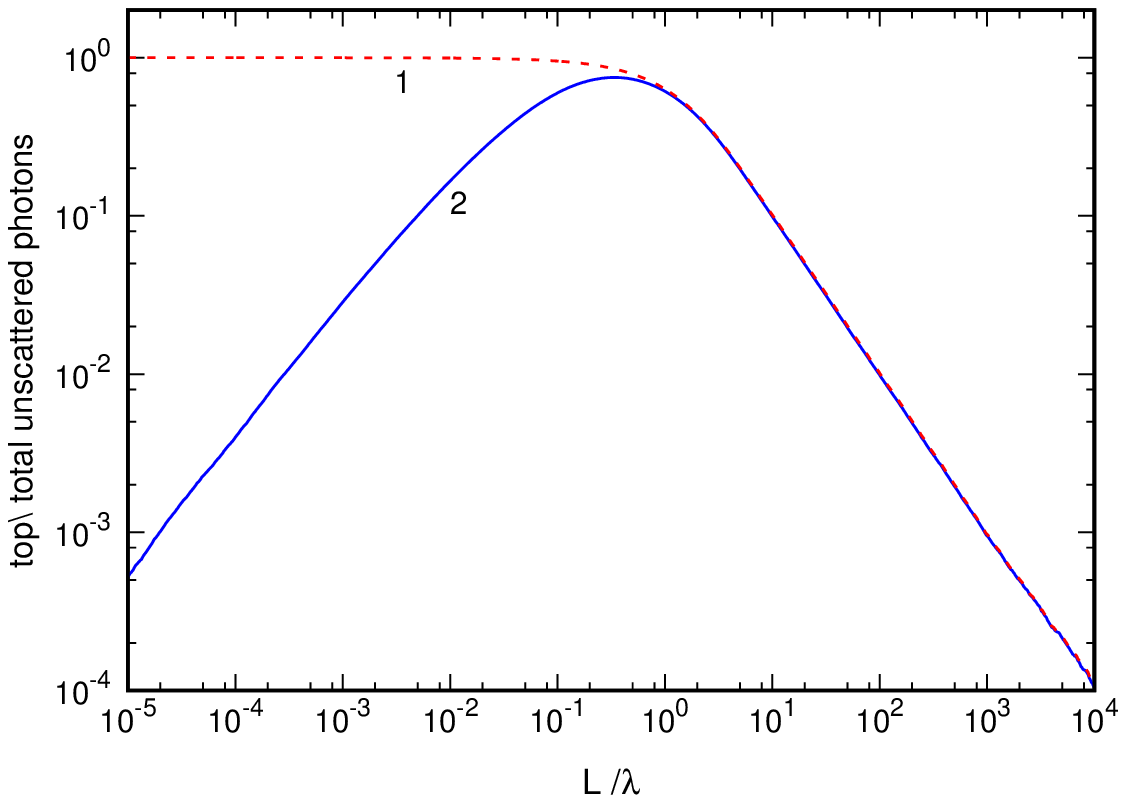}& \hspace{-0.6cm}
  \includegraphics[width=0.46\textwidth]{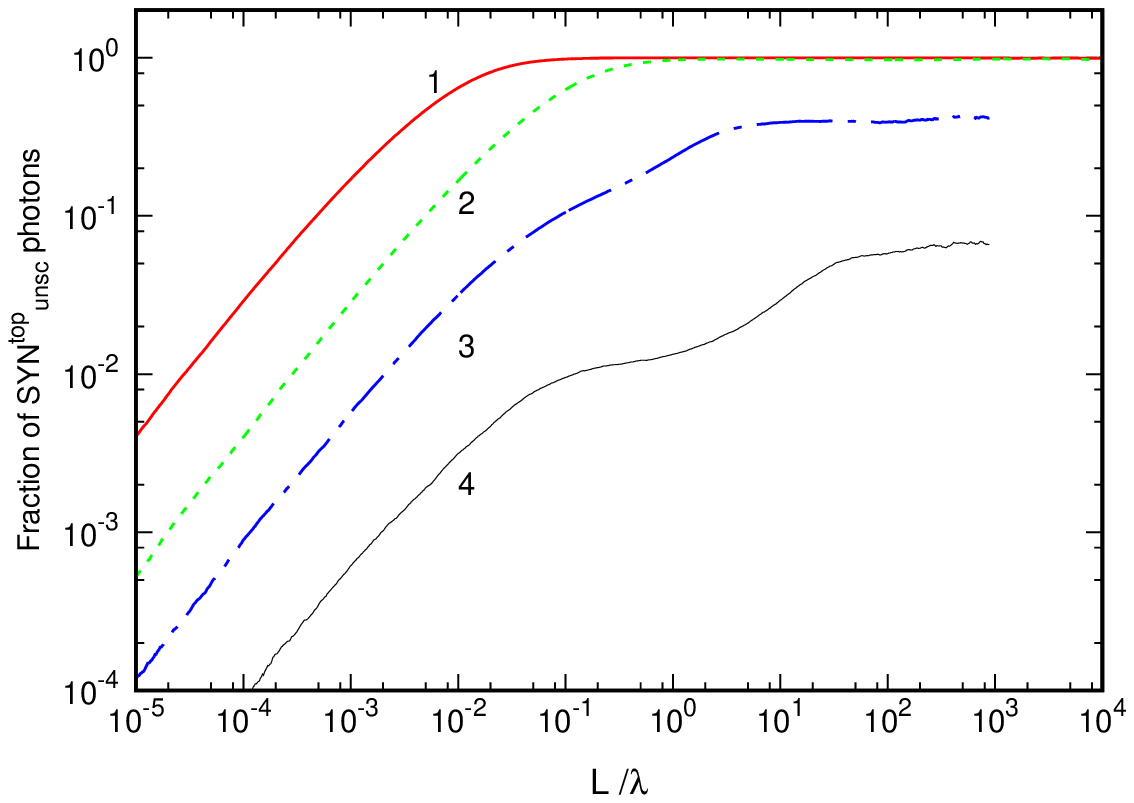}\\
\end{tabular}\vspace{-0.3cm}
\caption{{\it Left:} 
 Fraction of unscattered synchrotron photons to the total synchrotron photons for a jet apex angle $\theta_j$ = $5^\circ$
and radius R = 0.5L as a function of length of the emission region in units of scattering
mean free path. The solid curve represents the fraction of photons leaving from the
face of the cylinder towards the observer while the dotted curve represents the fraction of
total unscattered (top+sides) photons. {\it Right:} Fraction of unscattered
synchrotron photons leaving from the front face of the cylindrical emission region
for a fixed $\theta_j$ = 5 degree, but with different radius, here the curves 1, 2, 3 and 4 are for R = 5L, L/2, L/100 and L/1000 respectively. 
} 
\label{blz-ma1}
\end{figure}

The observed synchrotron emission is total synchrotron photons generated inside
the cylindrical emission region minus those that got IC scattered to higher energies.
Further, since synchrotron emission is mainly in forward direction 
along the jet axis, most of the emitted photons will exit from the face of the 
cylinder closer to the observer. Both of these depend on the geometry and
dimension of the emission region (scattering region) as shown in figure \ref{blz-ma1}
as well as the optical depth for SSC, which is measured along the length L. For a given $\lambda$, the unscattered photons existing from the 
front face will first increase and after reaching the maximum it will decrease
with
increasing L (as shown by curve 2 in left panel of figure \ref{blz-ma1}). It is a consequence of the conical direction of synchrotron emission, and of the increment of optical
depth (or average scattering number) by increasing L. 

In general, for a given $\theta_j$, mean free path of photon $\lambda$ and emission
region size L, the fraction of photons exiting from the front face of the cylindrical
emission region will decrease with decreasing R. Here, the unscattered synchrotron photons
exiting from the front of the cylindrical region ($SYN^{top}_{unsc}$)  are those
satisfying the condition z $>\rm L$ 
and $\rm\sqrt{x^2+y^2}$ $<$ 1.05R where (x,y,z) refer to spatial coordinate
of photons with centre of the bottom face as the origin. Particularly, for 10$\%$
of total synchrotron photons, to exit unscattered from the front face the length
of cylindrical region should be L = $5\times 10^{-5}\lambda$, 5$\times 10^{-4}\lambda$,
5$\times 10^{-3}\lambda$, $\sim \lambda$ for R = 50L, 5L, 0.5L, 0.01L respectively
(figure \ref{blz-ma1} right panel).
Further, we notice that 100$\%$ unscattered photon can escape from the 
top (i.e., $SYN^{top}_{unsc}$ = 100$\%$) when R $>$0.1L while for R$<$ 0.1L 
$SYN^{top}_{unsc}$ is always less than 100$\%$, e.g., for R=0.01L the maximum 
$SYN^{top}_{unsc}$ is 40$\%$ (as shown by curve 3 in right panel of figure 2).

\subsection{General properties of SED}\label{subsec:32}
As mentioned, in this section we are exploring the general properties of SED over issue, of limited energy range of synchrotron emission for calculating SSC (case 1), for subdominated random velocity components on IC spectra (case 2), and of dimension of emission region (case 3).

\subsubsection{Case 1}\label{subsec:321} 

\begin{figure*} 
\centering
\begin{tabular}{lcr}\hspace{-0.9cm}
  \includegraphics[width=0.36\textwidth]{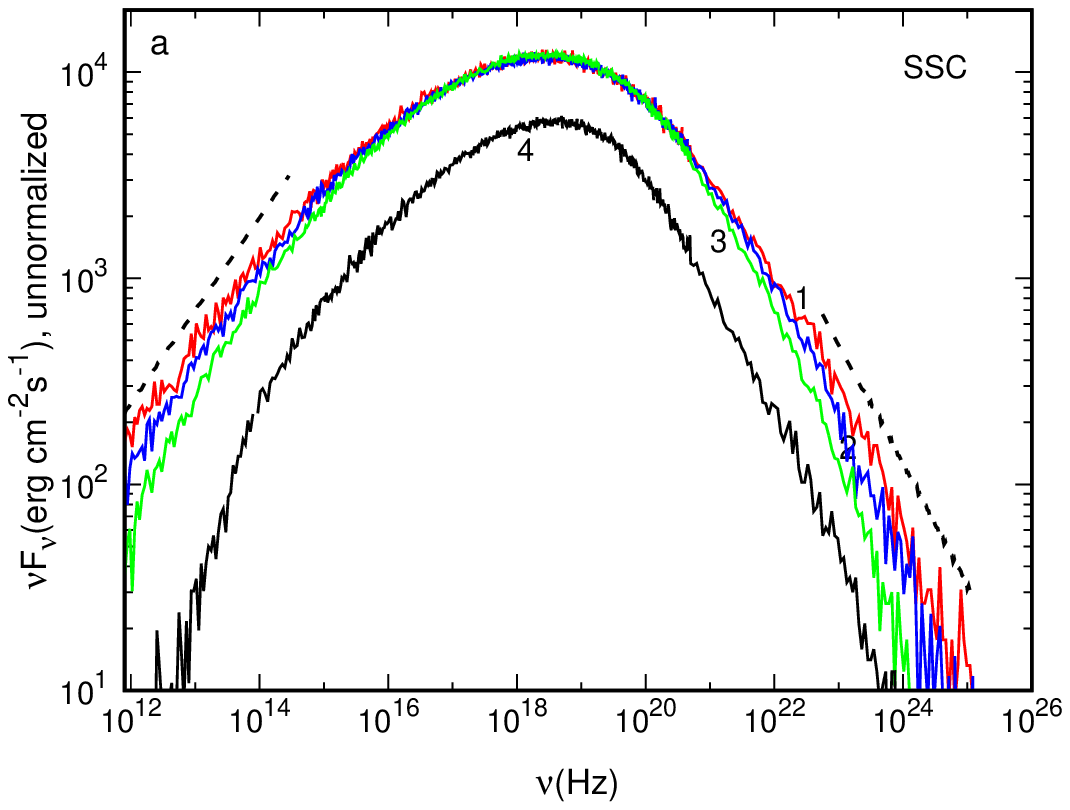}&\hspace{-0.90cm}
  \includegraphics[width=0.36\textwidth]{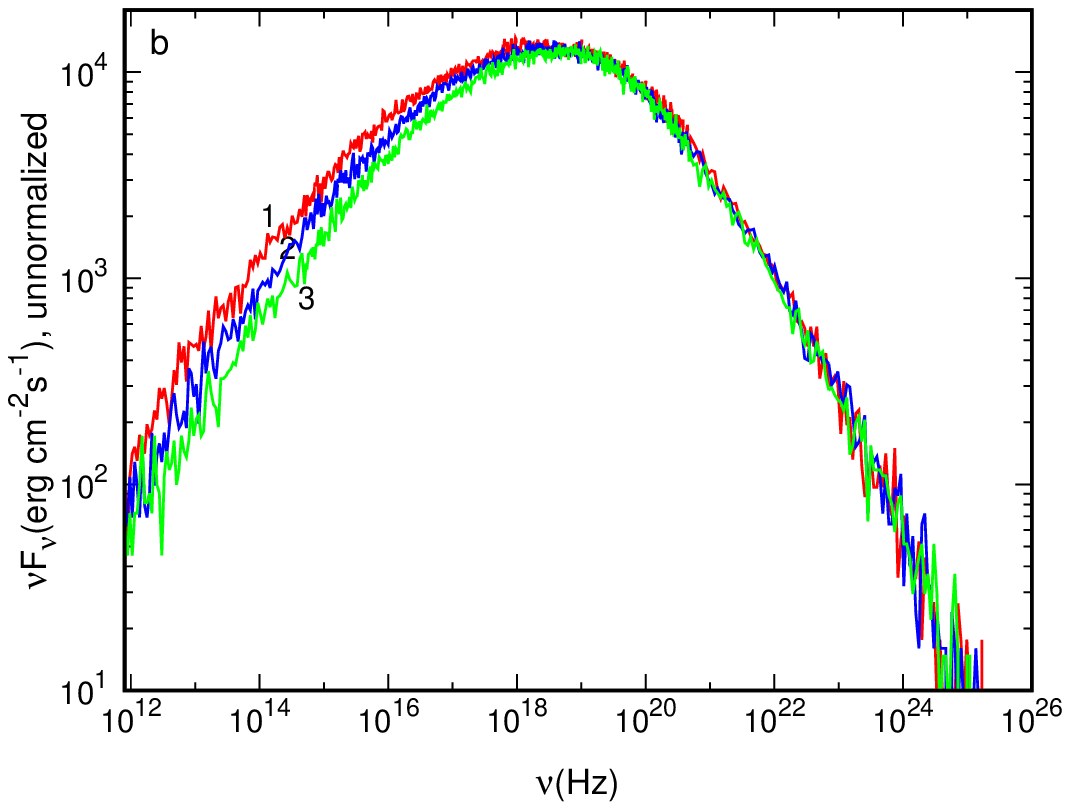}&\hspace{-0.90cm}
   \includegraphics[width=0.36\textwidth]{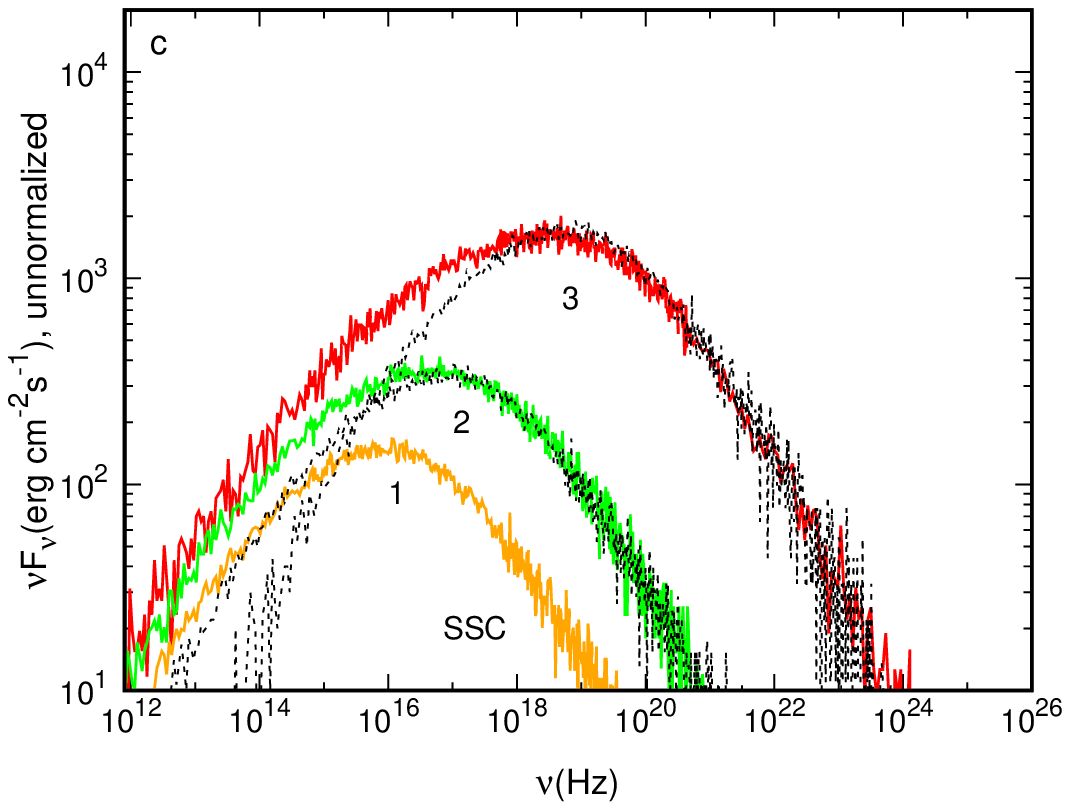}\\
\end{tabular}\vspace{-0.3cm}
\caption{The general behavior of SED when the $\nu^{syn}_{min}$ (or $B_o$) varies (panel (a))
and when the random component of electrons velocity $\gamma^{ran}$ varies (panel
(b) and (c)). In panel (b), the range of $\gamma^{ran}$ varies in such a way that average
of $\gamma^{ran}$, $\langle \gamma^{ran} \rangle$, is constant while in panel (c) $\langle
\gamma^{ran} \rangle$ also varies (see \S\ref{subsec:32}).}
\label{blz-gen}
\end{figure*}

In general, emission in radio bands below a few 10s of GHz is dominated by the
synchrotron-self absorption with emission from a much large region or multiple
emission regions, contrary to the rapidly variable NIR to $\gamma$-ray emission
-- the focus of our study. However, neglecting radio emission below 10 GHz can
affect the SSC as its the part of the seed photon for IC scattering. We, thus,
investigated this
effect on spectral shape of the IC component by limiting the synchrotron minimum
frequency $\nu^{syn}_{min}$ at different values. For this we fixed $\gamma_{min}$, the peak frequency of synchrotron spectra $\nu^{syn}_p$ and then compute the
axial magnetic field strength
$B_o$ for known $\nu^{syn}_{min}$
using  $\nu^{syn}_{min}$ =$\frac{\gamma_{min}^2 \langle \gamma^{ran}\rangle^2}{1+z}\nu_B$. 
We similarly compute the $\gamma_b$  using equation (\ref{syn-peak})  
for a given $\nu^{syn}_p$ and with the above estimated $B_o$.  Figure \ref{blz-gen}a shows the SED for three values of $\nu^{syn}_{min}$ = 10$^{10}$, 10$^{11}$, 10$^{12}$ Hz (curves 1, 2, and 3 respectively) with
$\gamma_{min}$ = 100, $\nu^{syn}_{p}$ = 10$^{14}$Hz, $\langle \gamma^{ran}\rangle$ = $\frac{\gamma_{min}}{2}$ = 50. For all the three cases the spectra
are similar, as expected, since the electrons energy distribution
are almost same but the seed photons for SSC (synchrotron photons) are differed
by $\nu^{syn}_{min}$ values while the spectral index $\alpha$ (where $\alpha$ = $(p-3)/2$ = 
-0.45, $\nu^{syn}$ $<$ 10$^{14}$Hz; and = $(q-3)/2$ = 0.55,  $\nu^{syn}$ $>$ 10$^{14}$Hz) 
is same.
Also, the spectral index of SSC is similar to the spectral index of synchrotron
emission, which is shown in figure \ref{blz-gen}a by plotting the dotted line at
lower ($\propto$ $\nu^{0.45}$) and higher ($\propto$ $\nu^{-0.55}$) frequency end of SSC.

The curve 4 of figure \ref{blz-gen}a is a SSC spectra when the seed photon is a
monochromatic photon of frequency 8$\times 10^{13}$ Hz with other parameters
being same. In this case also, the spectral index of SSC is similar to the spectral
index of the synchrotron emission. Hence, the spectral index of SSC spectra is
determined  mainly by the electron energy distribution. Further, we see that when the 
monochromatic seed photon frequency is around $\nu^{syn}_p$ (=10$^{14}$ Hz) then the 
SSC spectral peak frequency is same to the case where seed photon has a broken power
law with break frequency $\nu^{syn}_p$ (e.g., curve 1 of figure \ref{blz-gen}a) but
when it deviates much from $\nu^{syn}_p$ the corresponding SSC spectral peak frequency
also changes. Hence, the SSC spectral peak is mainly determined by the
spectral synchrotron peak frequency $\nu_p^{syn}$.

For the three $\nu_p^{syn}$ values mentioned above,
the estimated $B_o$ is 0.00014, 0.0014 and 0.014 gauss; $\sqrt{\langle \gamma^2 \rangle}$ is 1146, 681 and 404; $\gamma_b$
is 10000, 3162 and 1000 and $\langle \gamma \rangle$ is 431, 357 and 277 respectively. Hence,
the magnetic field strength determination by SED fitting
is degenerate with $\nu^{syn}_{min}$, when $\gamma_{min}$ and $\nu^{syn}_{p}$ are fixed.

%
\subsubsection{Case 2} \label{subsec:322}  
Next we studied the arbitrariness and effect of the low magnitude random component
of electrons velocity. 
We first
compute the SED by varying $\gamma_{min}^{ran}$, $\gamma_{max}^{ran}$
in such a way that $\langle \gamma^{ran} \rangle$ $\left(=
\frac{\gamma_{min}^{ran}+\gamma_{max}^{ran}}{2}\right)$ is fixed 
(shown in figure
\ref{blz-gen}b) and also explored the effect by varying the $\langle \gamma^{ran}
\rangle$ (figure \ref{blz-gen}c). Curves labeled 1, 2 and 3 in figure \ref{blz-gen}b
show the spectrum for a fixed $\langle \gamma^{ran} \rangle$ $\ \sim$ 50, but with 
$\gamma_{min}^{ran}$ = 1.8, 10 and 49, and $\gamma_{max}^{ran}$ = 100, 90 and 51 
respectively.
As can be seen, the low end of the SSC spectra are affected. In general, the slope of
the lower-end of spectra increase or spectra become harder with decreasing the separation between $\gamma_{min}^{ran}$
and $\gamma_{max}^{ran}$, achieving maximum hardness for $\gamma_{min}^{ran}$=$\gamma_{max}^{ran}$= $\langle \gamma^{ran} \rangle$. The same hold for EC as well.

Figure \ref{blz-gen}c
shows the SSC spectra for different $\langle \gamma^{ran}\rangle$. Curves 1, 2, and 3
correspond to $\langle \gamma^{ran}\rangle$ of 2, 6 and 51 with $\gamma_{min}^{ran}$ = 1.8
and $\gamma_{max}^{ran}$ = 2, 10 and 100 respectively. The peak of the curves 2 and 3 are shifted
by a factor 3$^2$ and 25.5$^2$ 
with respect to peak of curve 1, shown by overplotting 
curve 1 using this factor. In general, for given parameter sets, the peak of the SSC/EC
depends only on $\langle \gamma^{ran} \rangle$ 
and not explicitly on $\gamma_{min}^{ran}$, or $\gamma_{max}^{ran}$. The slope of the
lower-end of SSC/ EC spectrum, on the other hand, depends on $\gamma_{min}^{ran}$ and
$\gamma_{max}^{ran}$ with maximum for $\gamma_{min}^{ran}$=$\gamma_{max}^{ran}$=
$\langle \gamma^{ran} \rangle$.

\begin{figure*} 
\centering
\begin{tabular}{lcr}\hspace{-0.9cm}
  \includegraphics[width=0.36\textwidth]{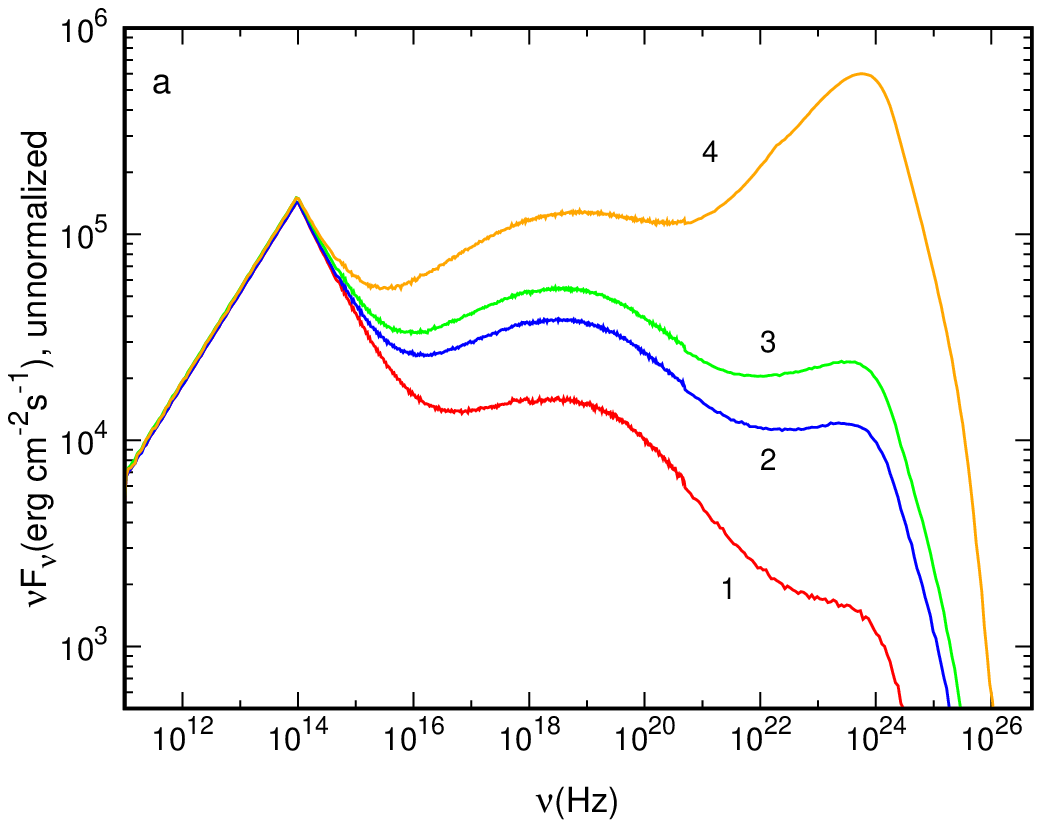}&\hspace{-0.90cm}
  \includegraphics[width=0.36\textwidth]{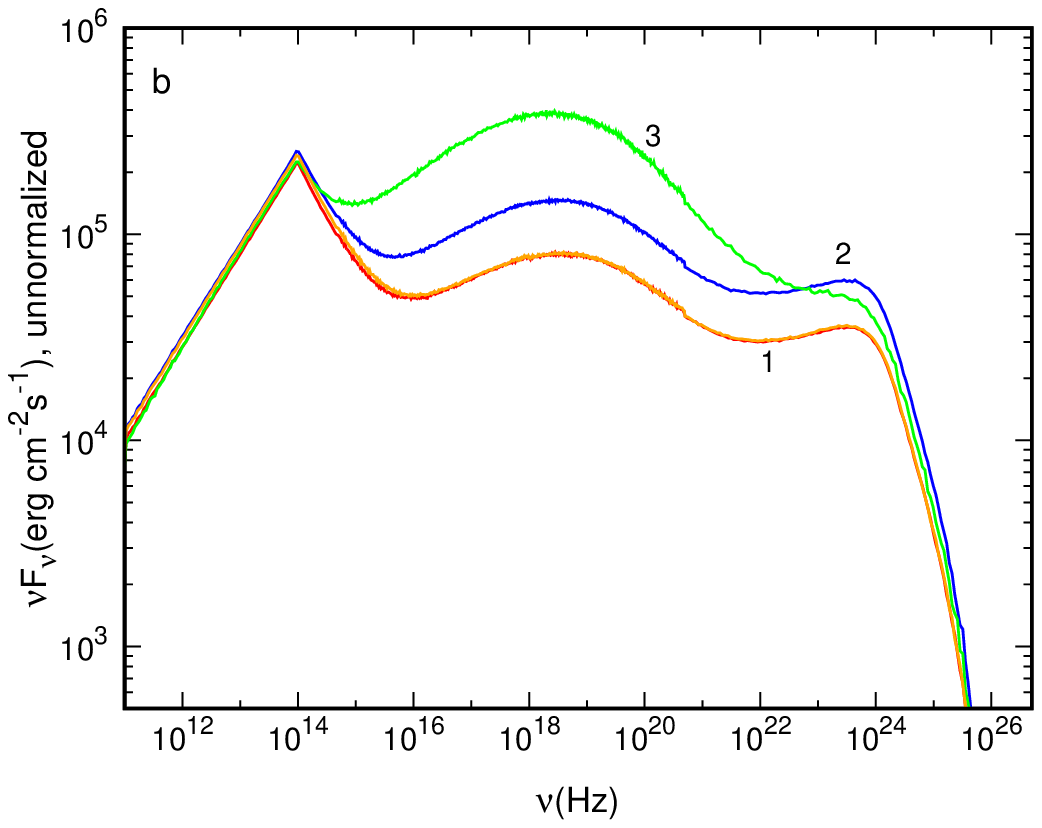}&\hspace{-0.90cm}
   \includegraphics[width=0.36\textwidth]{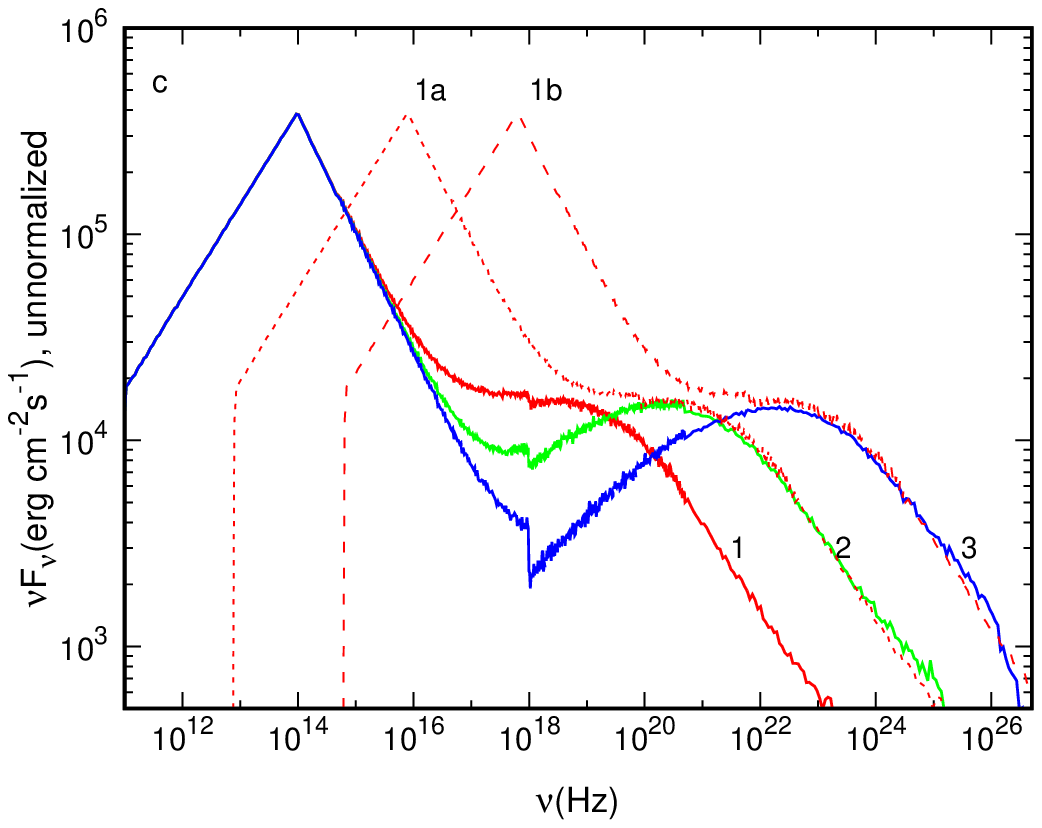}\\
\end{tabular} 
\caption{The general behavior of SED when L/$\lambda$ varies for fixed R/L (panel (a)),
when R/L varies for fixed L/$\lambda$ (panel (b)), and when $\gamma_{min}$ varies for
fixed L/$\lambda$, R/L (panel (c)). In panel (a), the curves 1, 2, 3, and 4 are for
L/$\lambda$ = 0.25, 0.5, 1 and 5 respectively at R/L=0.5. In panel (b), the curves 1,
2 and 3 are for R/L = 0.5, 0.05 and 0.005 respectively at L = $\lambda$. In panel (c),
the curves 1, 2 and 3 are for $\gamma_{min}$ = 100, 300 and 900 and curves 1a and 1b are
shifted curve 1 by factor 3$^4$ and 9$^4$ respectively.
} 
\label{blz-cylinder}
\end{figure*}

\subsubsection{Case 3} 
The other factor that affects the observed broadband spectrum is the dimension
of the emission region. In the present case of cylindrical geometry, its primarily
the ratio of radius to length, R/L and the ratio of length to the mean free path
of the photons, L/$\lambda$. Figures \ref{blz-cylinder}a and \ref{blz-cylinder}b show
the effect of L/$\lambda$ and R/L on the SSC spectra. As expected, the average
scattering number $\langle N_{sc} \rangle$ increases with L, so the maximum emission
of the SSC (photon counts corresponds to SSC peak) grows with L and for sufficiently
large value of L, a Wien's peak starts to appear (curve 4 of figure \ref{blz-cylinder}a). Though it
seems tempting that such peak may represent FSRQ SEDs, it corresponds to optically
thick regime, inconsistent with observations. In L $>$ $\lambda$ limit,
the ratio of emission at SSC peak to SYN peak increases with L, but is true only for
R/L $>$ 0.1 since all unscattered photons escape from the top (front face of the cylinder).
For R/L $<$ 0.1 case,
 $SYN_{unsc}^{top}$  is
less than 100 $\%$ and decreases with decreasing  R/L (right panel of figure \ref{blz-ma1}).
Hence, in latter case, for a given R/L ($<$0.1) the ratio of emission at
SSC peak to SYN peak remains
almost constant for any L value.

Figure \ref{blz-cylinder}b shows the effect of
variation of R for L = $\lambda$ and here $SYN^{top}_{unsc}$ =  97, 46 and
4.8 $\%$  for the curves 1, 2 and 3 respectively.
Effectively, $\langle N_{sc}
\rangle$ decreases with decreasing R/L, resulting in decrement of high SSC peak emission (which is appeared here around $10^{24}$ Hz) and
increment in ratio of emission at low SSC peak ($\sim \ 10^{19}$ Hz) to SYN peak.  
Figure \ref{blz-cylinder}c
shows the effect of variation of $\gamma_{min}$ only, as was done in figure \ref{blz-gen}c.
Here, curves 1, 2, and 3 are for $\gamma_{min}$ (the square root of $\langle
\gamma^2 \rangle$) of 100 (681), 300 (2045) and 900 (6137) respectively and
 correspondingly the SSC peak
frequency [$\nu_p^{ssc}$ $\propto$ $\langle \gamma^{ran} \rangle^2$ $\langle \gamma^2 \rangle^2$ $\nu_B$] of the curves 2 and 3 
increases by factor 9$^2$ and 81$^2$ with respect to the SSC peak frequency of the curve 1. Here, we like to stress that in our model the SSC peak is degenerate over $\langle \gamma^{ran} \rangle$ and $\gamma_{min}$ (or the square root of $\langle \gamma^2 \rangle$, or  $\langle \gamma \rangle$). The same SSC peak frequency can be modelled for lower $\langle \gamma^{ran} \rangle$ and correspondingly higher $\gamma_{min}$ (or higher anisotropicity in electron motion). 

\subsection{SED fitting Demonstration}\label{subsec:blazarSEDs}
Blazar characteristic broad double-humped SED is categorized primarily by the
frequency at which synchrotron emission peaks, and the ratio of
luminosities at high-energy hump peak to 
synchrotron peak in $\nu F_{\nu}$ representation; %
generally referred
in the literature as ``Compton dominance'' (CD) \cite[e.g.,][]{Dermer-Giebels2016}. Here, we introduce a general way of
SED modeling considering SSC for two cases, %
based on the Compton dominance. Case-I
considers blazars with CD $<$ 1 while Case-II with CD $\gtrsim$ 1. In general, the
 X-ray spectrum in case-II is comparatively harder
than the case-I. As shown in the previous section, the peak of SSC/EC depends only
on $\langle \gamma^{ran} \rangle$ while the slope of the low-energy end of SSC/EC
spectrum depends on $\gamma_{min}^{ran}$
and $\gamma_{max}^{ran}$ with slope achieving maximum for $\gamma^{ran}_{min} \approx \gamma^{ran}_{max}$ = $\langle \gamma^{ran} \rangle$ and minimum for $\gamma^{ran}_{min} \ll \gamma^{ran}_{max}$  $\approx 2\langle \gamma^{ran} \rangle$.  
So for SED modeling in the Case-II, we assumed $\gamma^{ran}_{min}
\approx \gamma^{ran}_{max}$ = $\langle \gamma^{ran} \rangle $  while $\gamma^{ran}_{min}$=1.8, $\gamma^{ran}_{max}$ = $\gamma_{min}$ for Case-I. Below, we demonstrate our approach without 
bothering about the fairness of parameters vis-a-vis observations.

\begin{figure*} 
\centering
\begin{tabular}{lcr} \hspace{-0.9cm}
  \includegraphics[width=0.36\textwidth]{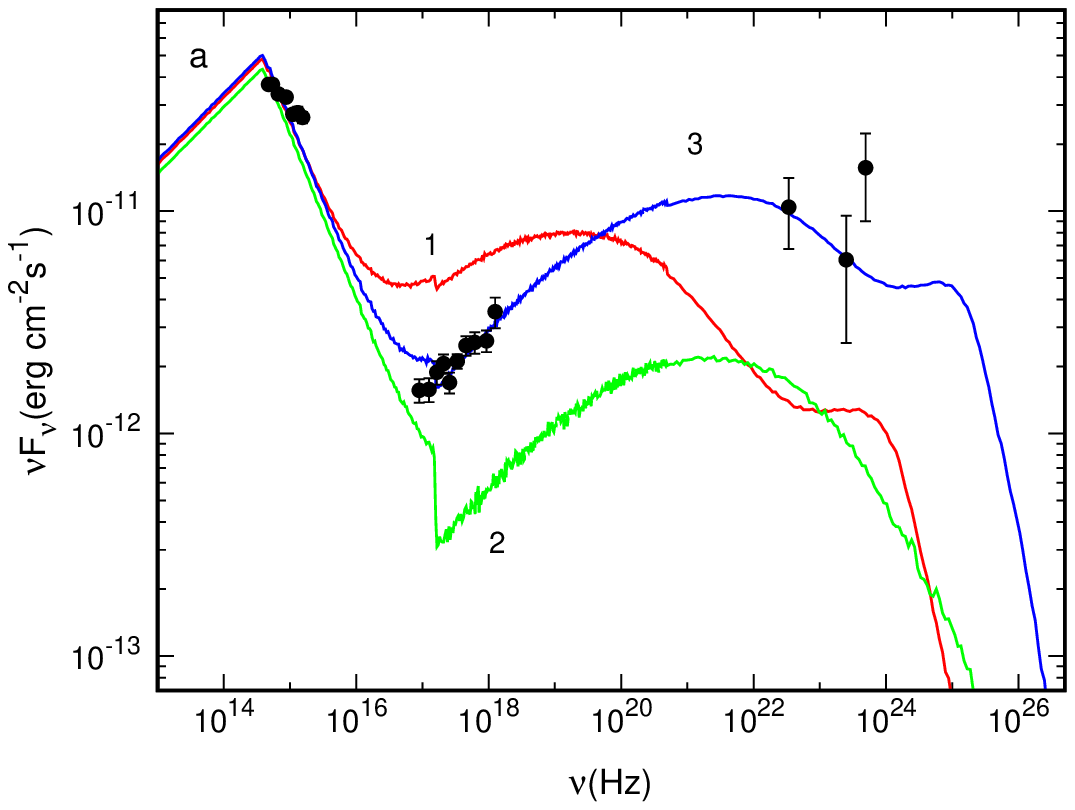}&\hspace{-0.90cm}
  \includegraphics[width=0.36\textwidth]{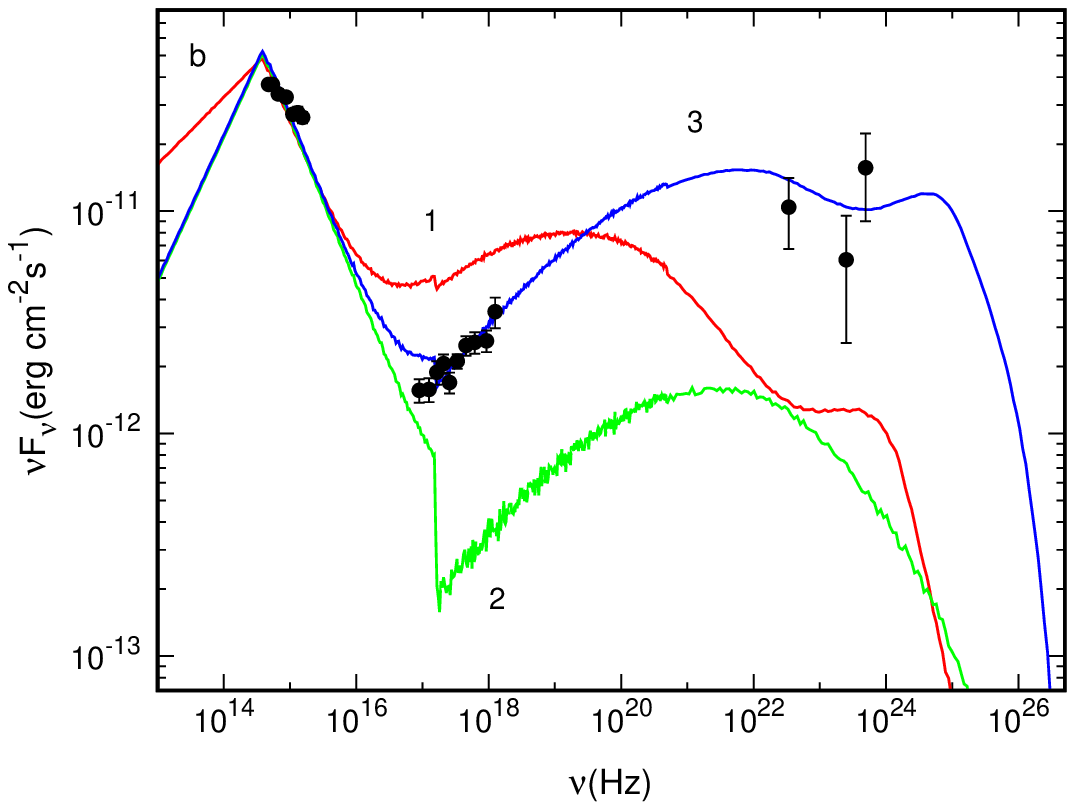}&\hspace{-0.90cm}
   \includegraphics[width=0.36\textwidth]{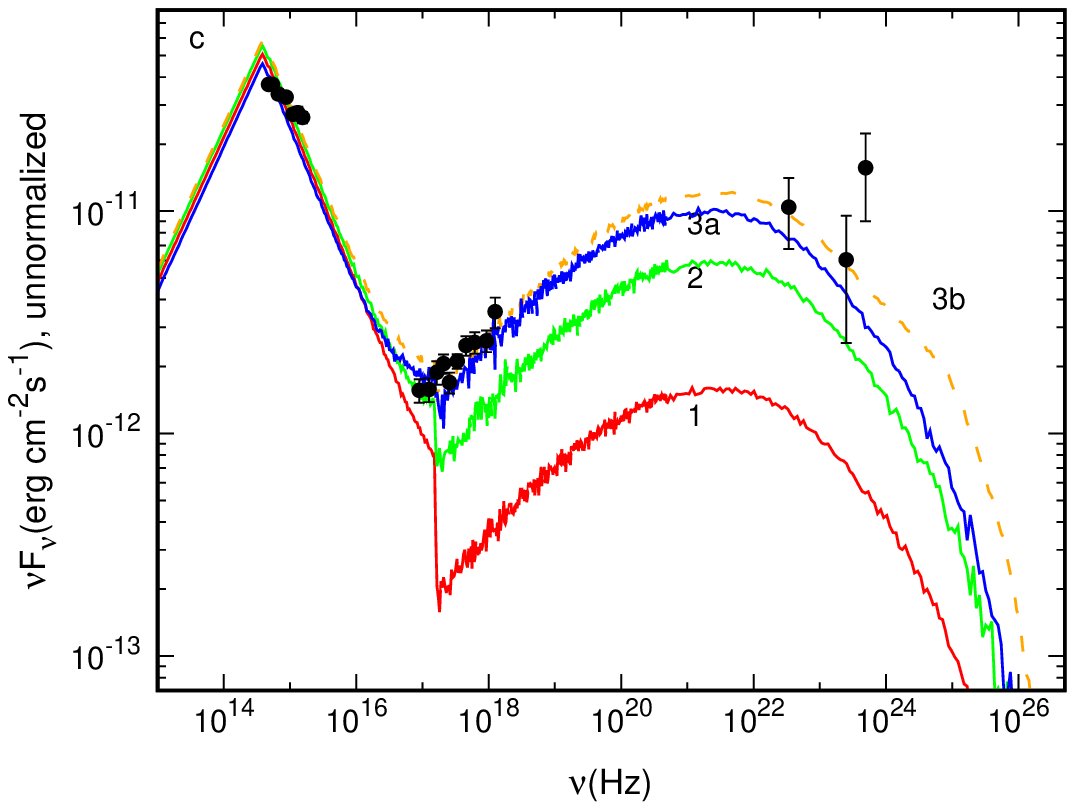}\\
\end{tabular} 
\caption{ A Demonstration of SED modeling by SSC process only for  CD $<$ 1, the data points are for quiescent state of OJ 287 \citep{Kushwaha-etal2018a}.
  In all three panels, the curve 1 is a starting/trial curve (except panel (c) where curve 1 is same as curve 2 of panel (b)). And curve 3 is a best modeled curve. The best modeled curve has been obtained in panel (a), (b), (c) by changing the combination of $\gamma_{min}$ $\&$ L; $p$, $\gamma_{min}$ $\&$ L; and R/L, $p$, $\gamma_{min}$ $\&$ L respectively.
} 
\label{blz-demo-I}
\end{figure*}
\subsubsection{Case-I SED}
Figure \ref{blz-demo-I} demonstrates the SED modeling by SSC process for case-I.
We have chosen the observation ID of source OJ 287 \cite[from][]{Kushwaha-etal2018a}
where CD  $\approx$ 1/4. 
We have attempted to model the SED in three different ways (as shown in figure \ref{blz-demo-I}
(a),(b),(c)) exploiting the degeneracies between parameters. For all the cases, the starting parameters set is $\gamma_{min}$ =100 and
L = 1.1$\times$10$^{14}$cm = 0.35$\lambda$, $p$ = 2.4,  R/L=0.5; $q$= 4.5, $\gamma_{max}=40000=20
\gamma_b$, $\langle \gamma\rangle$ = 260, $B_o$=0.014 gauss, $\theta_j$=10 degree,  $n_e$ = 10$^{14}$ cm$^{-3}$ (here $\langle \gamma \rangle / \langle \gamma^{ran} \rangle $ = 5.2, meant an anisotropic flow),
the corresponding spectrum is shown by curve 1 in panel (a) and
(b), as a trial or guess
  values. In which the SSC peak is around 10$^{19}$ Hz and the intensity of minimum X-ray frequency is higher than the observed one. 
To decrease the intensity at the lower-end of
X-ray spectrum and increase the SSC peak frequency from curve 1, in panel (a), we had increased the $\gamma_{min}$ to 350 (shown by curve 2).
In panel (b) we had decreased the p to 1.7 and increased the $\gamma_{min}$ to 250 (shown by curve 2), resulting in a shift of SSC peak frequency to 10$^{21}$ Hz. The best
model description are shown by curve 3 in panels (a) and (b), where $\gamma$-ray peak
intensity is matched by increasing the size of region (results in $\langle N_{sc}
\rangle$ increment) with L =2.45$\times$10$^{15}$, 7.52$\times$10$^{15}$ cm for 
panel (a) and (b) respectively.

In panel (c), we start with the curve 2 of panel (b) but
now to increase the ratio of $\gamma$-ray peak to synchrotron peak intensity (as shown in figure \ref{blz-cylinder}b)
we decreased the R/L to 0.1 (curve 2) and to 0.03 (curve 3a, which is also a best modeled curve with L =1.1$\times$10$^{14}$ cm).
And the another best model description curve 3b  has been obtained from curve 2 by increasing
 L to 1.1$\times$10$^{15}$cm. $\langle N_{sc} \rangle$ for best model description
curve of panels (a), (b) and (c) (curves 3a and 3b) are 1.12, 1.25, 1.003 and 1.04, and 
$SYN_{unsc}^{top}$ is 98, 98, 1.1 and 20 $\%$ respectively.  
In short, we have modeled the same data in 3 different way using degeneracy between
parameters as discussed in the previous subsections. In first approach, we changed only
$\gamma_{min}$ and L while in second, we decreased the spectral index $p$ and comparatively
lowered  $\gamma_{min}$ but with a higher L than the first approach. We
also found that of all, in third approach, by changing R/L which leads to lowest L.%

\begin{figure} 
\centering
\begin{tabular}{lr}\hspace{-.8cm}
  \includegraphics[width=0.36\textwidth]{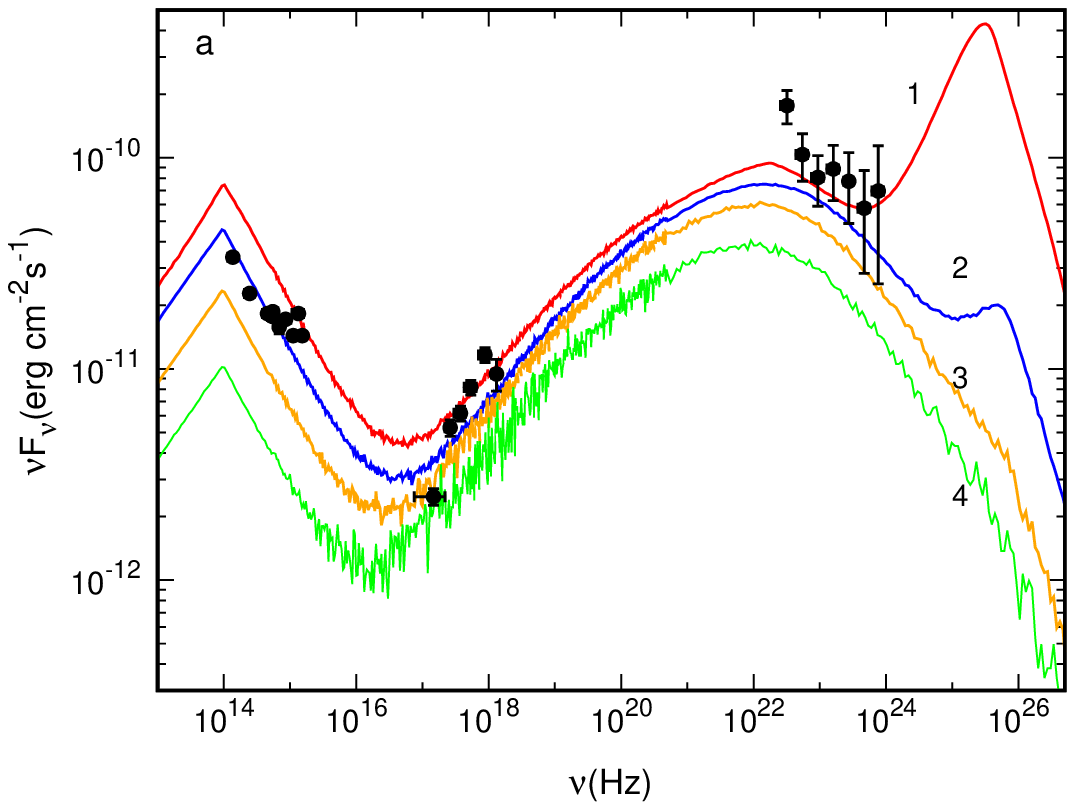}&\hspace{-0.80cm}
  \includegraphics[width=0.36\textwidth]{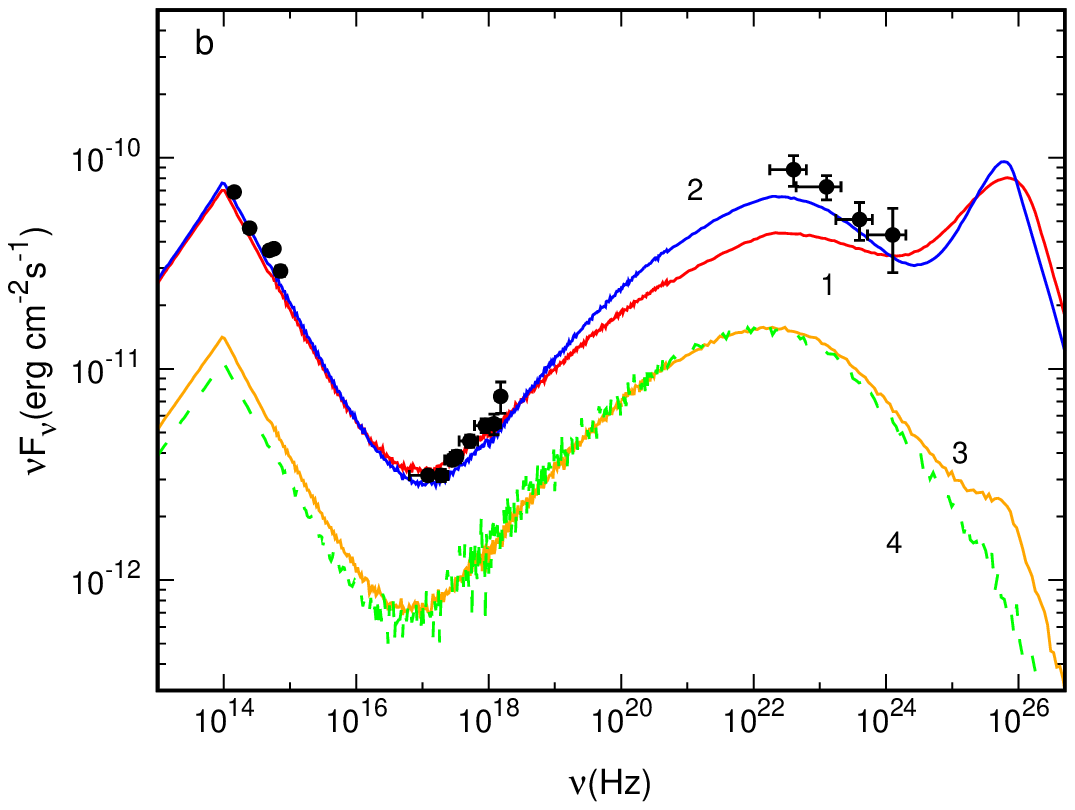}\\ 
\end{tabular} 
\caption{A Demonstration of SED modeling by SSC process only for CD $\gtrsim$ 1. The panel (a) is for quiescent phase of 3C 454.3 \citep{Shah-etal2017} and (b) is for flaring phase of OJ 287 \citep{Kushwaha-etal2013}, here all curves of both panels are best modeled one which is obtained mainly by varying R/L. In panel (a) the curves 3 and 4 are lowered by factor 2 and 4 respectively and in panel (b) the curves 3 and 4 both are lowered by factor 5.
} 
\label{blz-demo-II}
\end{figure}        

\subsubsection{Case-II SED}
Next we demonstrate the SED fitting by SSC only for case-II. For this, we
have selected two observation IDs, a quiescent phase of 3C 454.3 (FSRQ, \citealp[][]
{Shah-etal2017}) and a flaring state of OJ 287 (BL Lacs object, \citealp[][]{Kushwaha-etal2013}).
CD is $\sim$ 2 and
$\sim$ 1, shown respectively in figure \ref{blz-demo-II}a and \ref{blz-demo-II}b. %
For SED modeling,
we have used the random component of electron $\gamma^{ran}_{min}$ $\approx$ $\gamma^{ran}_{max}$ = $\langle \gamma^{ran} \rangle$ for panel (a), and $\gamma^{ran}_{min}$=1.8, $\gamma^{ran}_{max}$
= $\gamma_{min}$ %
for panel (b). For $\gamma$-ray dominated emission, the model requires
either a large average scattering (or L/$\lambda >$ 1, so top-unscattered SYN photon
will be less), or small R/L value (so again top-unscattered SYN photon will be less,
see figure \ref{blz-cylinder}b). 
Mainly, we have attempted by decreasing  R/L
value in order to get a small average scattering number. Since in large 
scattering limit, we will see Wien peak like features (shown by curve 1 and 2 
in both panels of figure 6) and also one can not avoid the pair-production over IC process. To best of our knowledge, none of the observed SED had reported Wien peak feature seen in the model curves at the high energy ends.

In both panels \ref{blz-demo-II}a and \ref{blz-demo-II}b, all curves are best
description to the data points except that some are shifted down from the data points
for clarity. Here, curves 1, 2, 3 of figure \ref{blz-demo-II}a and 2, 3, 4 of figure \ref{blz-demo-II}b
are for R/L = 0.05, 0.005, 0.0005 respectively while R/L = 0.00005 for curve 4 (in figure \ref{blz-demo-II}a) and
R/L = 0.5 for curve 1 (in figure \ref{blz-demo-II}b). L values for figure \ref{blz-demo-II}a curves 1, 2, 3 and 4
are 1.0 $\times 10^{17}$ ($\sim$4$\lambda$), 2.8 $\times 10^{16}$ ($\sim$0.6$\lambda$),
1.8 $\times 10^{16}$ ($\sim$0.4$\lambda$) and 1.8 $\times 10^{16}$($\sim$0.4$\lambda$) cm,
 $SYN_{unsc}^{top}$ is 69, 4, 0.3 and 0.03 $\%$, and $\langle N_{sc} \rangle $ is 2.04, 1.06, 1.007 and 1.001 respectively. Similarly,  L values
for curves 1, 2, 3, and 4 of figure \ref{blz-demo-II}b are 1.3 $\times 10^{17}$
($\sim$1.9$\lambda$), 7.9
$\times 10^{16}$ ($\sim$1.5$\lambda$), 1.3$\times 10^{16}$($\sim$0.25$\lambda$) and 5.3$\times 10^{15}$($\sim$0.1$\lambda$)  cm; and  
$SYN_{unsc}^{top}$ is 99, 55, 2.5 and 0.14 $\%$ and $\langle N_{sc} \rangle $ is
1.55, 1.37, 1.02 and 1.002 respectively.  In short, to model the Compton dominated SEDs in single scattering limit in SSC process, 
   the lowest R/L value ($<$ 0.01) is more preferable.
In both cases of SED modeling demonstration, 
 we have fixed an electron density, $n_e$ = $10^{14}$ cm$^{-3}$ which correspond to
the mean free path $\lambda$ = 4$\times 10^{14}$ cm for $\langle \gamma \rangle$ = 260. Here, we like to stress that for a comparatively smaller $n_e$ ($\sim 10^8$  cm$^{-3}$) the size L (for same SEDs) will become an order of 100 pc, as now $\lambda$ = 4$\times 10^{20}$ cm with same $\langle \gamma \rangle$. In this extreme case, one can not assume a cylindrical emission region, and in SED calulations one have to take an account of the density variation with length L, since for conical region the density decreases by square of the
radius of cross section.
  
\section{SED Comparison and Discussions}\label{sec:SEDcomp}
We performed a comparative study of blazars SED modeling from a local relativistic
anisotropic flow in a conical jet scenario and explored the physical parameters space.
We modeled five different SEDs of blazars, three for BL Lac
object, one each for OJ 287 (z = 0.306), S5 0716+714 (z = 0.31) and PKS 2155-304 (z = 0.116),
and the rest two for FSRQ 3C 454.3 (z = 0.859). The BLL OJ 287 has exhibited diverse
spectral phases, from LBL \cite[][]{Kushwaha-etal2013} to LBL+HBL \cite[][]{Kushwaha-etal2018b,Kushwaha-etal2018a}. For BLLs broadband SEDs, we have taken LBL phase of
OJ 287, IBL S5 0716+714 \cite[from][]{Chandra-etal2015}, and HBL PKS 2155-304 \cite[from][epoch 2012]{Gaur-etal2017}. For FSRQ, we have selected a quiescent and a flaring
phase of  3C 454.3 from \cite{Shah-etal2017}. In our conical jet model for a given
$\gamma$-ray flare time scale ($t_{var}$) the emission region size
L can be estimated as, 
\begin{equation}
 {\rm L} \leq \frac{c t_{var}}{1+z}.
\end{equation}
We defined, $L^o$ = $c t_{var}/(1+z) $ as a maximal observable emission size. For flaring phase of 3C 454.3, 
\citet[][]{Shah-etal2017} have reported a 10 hours $\gamma$-ray
variability, so L $\leq$ 5 $\times$ 10$^{14}$ cm. In general, for FSRQ SEDs, in addition
generally to agreed EC scenario, we have also attempted to model the SED by SSC
only. For BLLs, however, we have considered SSC only. For all four sources,
  we consider an average semi-aperture angle $\theta_j \sim$ 5 degree \cite[like,][]{Ghisellini-2009}.
In section  \S\ref{subsec:32}, we have shown that the magnetic field strength is degenerate with $\nu^{syn}_{min}$ for fixed $\gamma_{min}$ and $\nu_p^{syn}$ for given SED. We have computed the magnetic field strength $B_o$ by fixing the $\nu^{syn}_{min}$ as, $\nu_{min}^{syn}$ = $\frac{\gamma_{min}^4}{4(1+z)}\nu_B$, here $\gamma^{ran} = \gamma_{min}/2$ (e.g., see equation (\ref{syn-peak})).
Like previous section, we fix the electron density $n^*_e$ =$10^{16}$ cm$^{-3}$ for all Obs. ID, this will provide a trial L to start with. 
Later, we
  make this L to order of $L^o$ by adjusting (increasing or decreasing) the  $n^*_e$ by factor $a$ as, $n_e = a\ n^*_e$; where $a$ = L $/ L^o$. Since, in IC process the spectrum depends on average number scattering $\langle N_{sc} \rangle$, it degenerates over $n_e$ (or $\lambda$) and L. Mainly, we are estimating the electron density for given observed SEDs and its variability time scale (or $L^o$). In additions, the model curves are well describe the observed data points, when it is computed in single 
  scattering limits, $\langle N_{sc}\rangle \sim$1 (i.e., optically thin medium).  

\subsection{Considering both SSC and EC}\label{subsec:ECdom}
The broadband SEDs of FSRQs are usually modeled by SYN, SSC, and EC with EC component
dominating over SSC in general. The generally invoked seed photon source for EC
is torus (IR) and broad line region (BLR). Here, we considered these two, 
described by a blackbody profile with temperature corresponding to the frequency at which their emission peaks. The temperature for torus is $\leq$ 1800
K while BLR  is $\sim$ 40000 K. We have chosen two temperature values ($T_{EC}$),
500 and 50k K to extract a general picture in order to understand which describes
the data better assuming simply an isotropic distribution of seed photons inside
the medium.

\begin{figure*} 
\centering
\begin{tabular}{lr}\hspace{-.28cm}
\includegraphics[width=0.46\textwidth]{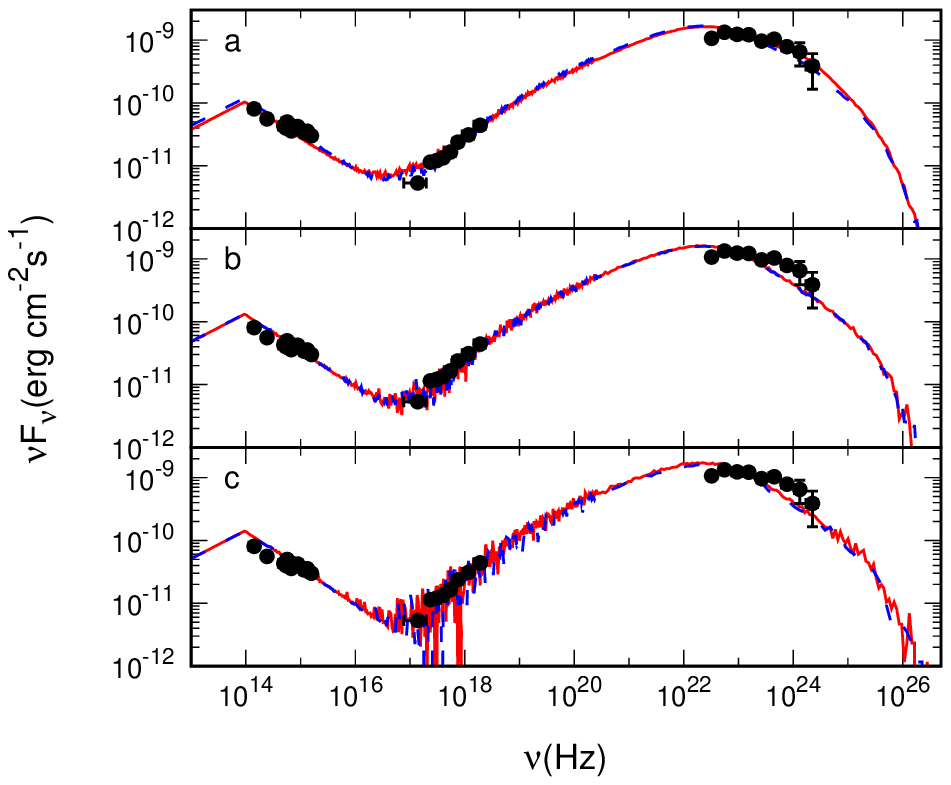}&\hspace{-1.20cm}
\includegraphics[width=0.46\textwidth]{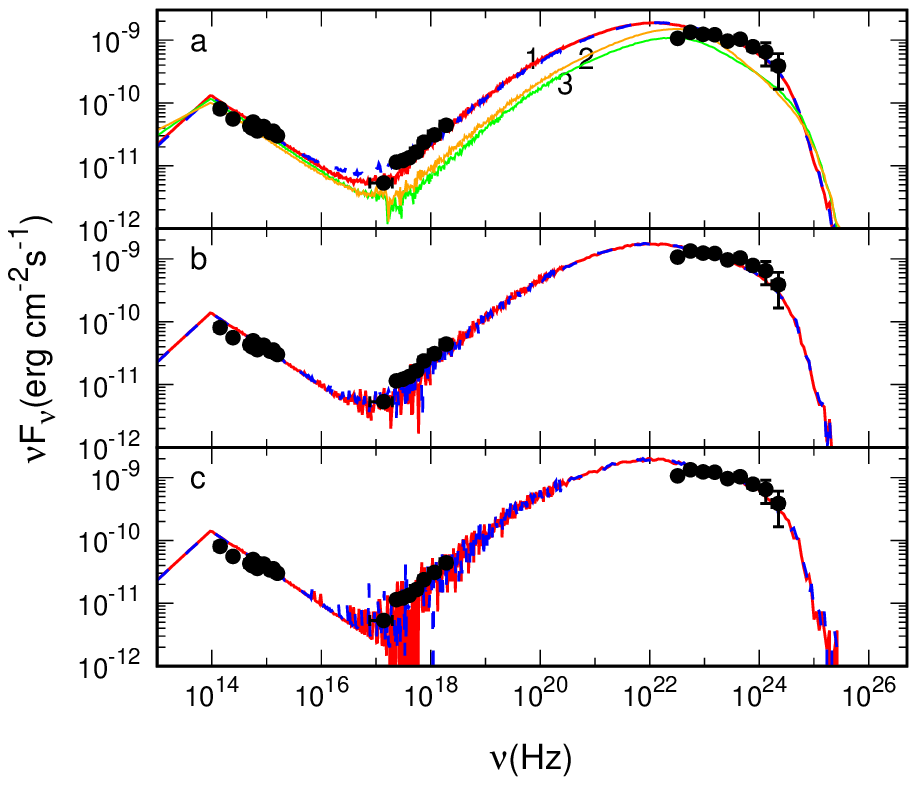}\\
\end{tabular}\vspace{-0.3cm}
\caption{SSC+EC explanation of high energy emission of flaring phase of 3C 454.3 \citep{Shah-etal2017}. The EC blackbody temperature is 500  and 50k K for left and right panel respectively. 
The panels (a), (b) and (c) are for $SYN^{top}_{unsc}$ 100, 10, 1 $\%$ respectively, computed simply by decreasing L from (a) to (c) (see \S3.1). The curves are total emission under synchrotron, SSC and EC explanation in which the dotted curve is for R/L = 5 and solid cure is for R/L = 0.5. 
} 
\label{blz-fl-sed}
\end{figure*}

\begin{figure} 
\centering
\begin{tabular}{lr}\hspace{-1.3cm}
  \includegraphics[width=0.46\textwidth]{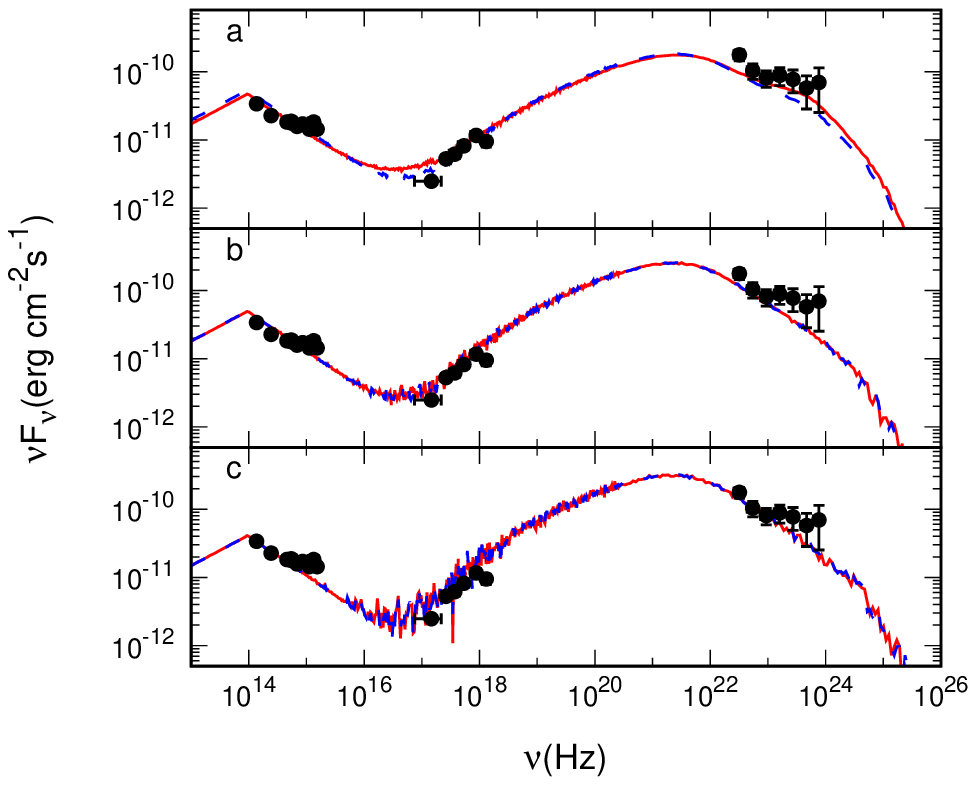}&\hspace{-1.70cm}
  \includegraphics[width=0.46\textwidth]{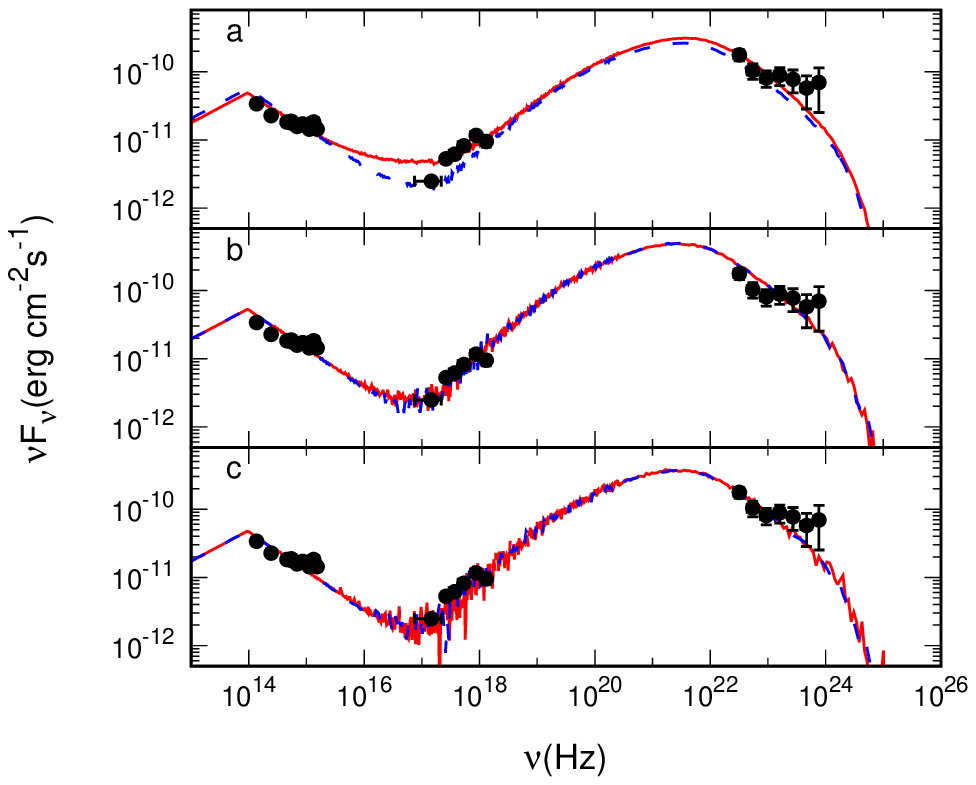}\\
\end{tabular}\vspace{-0.3cm}
\caption{SSC+EC explanation of high energy emission of quiescent phase of 3C 454.3 \citep{Shah-etal2017}, the rest is same as in figure \ref{blz-fl-sed}.
} 
\label{blz-qs-sed}
\end{figure}

\begin{figure} 
\centering
\begin{tabular}{l}\hspace{-0.8cm}
\includegraphics[scale=0.56]{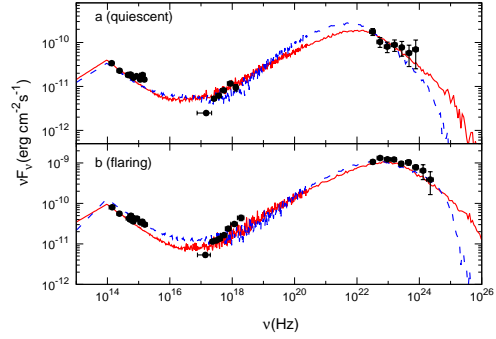}
\end{tabular}\vspace{-0.3cm}
\caption{SSC+EC explanation of high energy emission of  quiescent (upper row) flaring (lower row) phase of 3C 454.3 \citep{Shah-etal2017}. In both panel, $SYN_{unsc}^{top}$ = 1 $\%$, R/L = 0.01; and T$_{EC}$ = 500, 50k K for solid and dotted curve respectively.
 }
\label{blz-qs-flr001}
\end{figure}

Figures \ref{blz-fl-sed} and \ref{blz-qs-sed} show modeling of a
flaring and a quiescent phase of FSRQ 3C 454.3. In both the figures, the left and right panels 
are for 500 and
50k K with panels (a), (b), (c) corresponding to $SYN_{unsc}^{top}$ of $\sim$100,
10 and 1 $\%$ respectively, computed simply by decreasing L while rest
parameters (as shown in table 1) being fixed. As discussed in section \S3.1 the $SYN_{unsc}^{top}$ is 100, 10 and 1 $\%$ for L/$\lambda$ = 0.03,
5.1$\times 10^{-4}$, 2.9$\times 10^{-5}$ at R/L = 5; and = 0.3, 5.1$\times 10^{-3}$, 2.9$\times 10^{-4}$ at R/L = 0.5 respectively, 
L is decreasing from panel (a) to (c). The solid curve
is for R/L = 0.5 and dotted curve is for R/L = 5 with same parameters. 
Both curves are almost identical, since we are changing only geometry parameters R and L in such a way that $SYN_{unsc}^{top}$ is constant, and as their ratio R/L is greater than 0.1 so average scattering number $\langle N_{sc} \rangle$ is also similar. When R $<$ 0.1L, as discussed in previous section, $SYN_{unsc}^{top}$ never reaches
100$\%$ and the difference in SSC/EC spectra is sent to appear. In figure \ref{blz-qs-flr001} we have modeled the quiescent and flaring phase of 3C 454.3 for R/L = 0.01   which corresponds to $SYN_{unsc}^{top}$  $\sim$1$\%$.

\begin{table*} 
\centering
\caption{Spectral free  parameters for SED modeling (SYN+SSC+EC) of FSRQ 3C 454.3 as shown in figures \ref{blz-fl-sed}- \ref{blz-qs-flr001}. 
$n_e$ is estimated as, $n_e$ = $a\ n^*_e$;  where, $a$ = L$/L^o$, $L^o \sim 5\times10^{14}$ cm, and L is computed for $n^*_e$ = $10^{16}$ cm$^{-3}$ (see text for details).}
\begin{footnotesize}
\begin{tabular}{|p{1.45cm}*{10}{p{0.85cm}}| }
  \hline
  & \multicolumn{4}{|c|}{figure \ref{blz-fl-sed} (flaring)} & \multicolumn{2}{c|}{figure \ref{blz-qs-sed}  } & \multicolumn{2}{c|}{figure \ref{blz-qs-flr001} } & \multicolumn{2}{c|}{figure \ref{blz-qs-flr001} } \\ \cline{2-5}
 & \multicolumn{1}{|c}{} &  \multicolumn{3}{|c|}{(a)$_r$, for curves} &   \multicolumn{2}{c|}{(quiescent)} & \multicolumn{2}{c|}{(quiescent)} & \multicolumn{2}{c|}{(flaring)}\\ \cline{3-5}
  & \multicolumn{1}{|l|}{(a)$_l$}  &  1 &  2 & \multicolumn{1}{l|}{ 3} & (a)$_l$  & \multicolumn{1}{l|}{(a)$_r$} & (a)$_S$ & \multicolumn{1}{l|}{(a)$_D$} & (b)$_S$ & \multicolumn{1}{l|}{(b)$_D$} \\\cline{2-5} \cline{6-7} \cline{8-11}
$\langle N_{sc} \rangle$   & 1.05 & 1.03  &1.03  &1.03  & 1.05 & 1.04 & 1.0 & 1.0 & 1.0   & 1.0  \\
$\lambda$ (10$^{16}$cm)  & 13.0 & 0.18 & 0.26  & 0.18   & 1.3   & 0.016 & 5.9 & 0.031 & 48.0  &  0.76   \\
  L$^\#$(10$^{14}$cm) & 39.0  & 0.54  & 0.78  & 0.54   & 3.9   & 0.048 & 10.1  & 0.043 & 86.4  & 1.0    \\  
  $B_o^{\#\#}$ (G) & 2.1e-4 & 1.0e-2  & 2.2e-2 & 1.6e-2    & 2.1e-3 & 1.7e-1 & 4.9e-4 & 1.0e-1 & 6.1e-5 &  2.2e-2   \\
  T$_{EC}$(K) &500  &50k  & 50k  &50k     &500 & 50k &500 & 50k &  500  &  50k \\
  $\langle \gamma \rangle^{**}$ & 604  & 321   & 270   &  207  & 340 & 113 & 491 & 132 & 831 & 459  \\
  $\gamma_b^{**}$ & 6900  & 1509 & 2156 & 2372  & 3881 & 1293 & 5606 & 1509 & 9488 & 2156  \\
  $\gamma_{min}^{**}$ & 160 & 35 & 50 & 55  & 90    & 30 & 130 & 35 & 220 & 50   \\
  $p$ & 2.1 & 1.4  & 1.8 & 2.1  &2.1 & 2.1 & 2.1 &2.1 & 2.1 & 1.4   \\
  $q$ & 4.1 & 4.1 & 4.1 &  4.1  &4.1 & 4.1 & 4.1 &4.1 & 4.1  & 4.1 \\
  R/L & 5$^{***}$   &5$^{***}$   & 5$^{***}$ & 5$^{***}$   & 5$^{***}$   & 5$^{***}$   & 0.01 & 0.01 & 0.01  & 0.01 \\\hline
  \multicolumn{11}{l}{$^{**}$ in units of $m_ec^2$; $^{***}$ $\sim$100 $\%$ unscattered synchrotron photons escape from top of the emission region}\\
  \multicolumn{11}{l}{$^\#$ L = 5.1$\times 10^{-4}$, 2.9 $\times 10^{-5}$ $\lambda$ for  $SYN^{top}_{unsc}$ = 10, 1 $\%$ when R/L =5 and so on (see text). }\\
   \multicolumn{11}{l}{ $^{\#\#}$ $B_o$ has been computed for fixed $\nu_{min}^{syn}$ as, $\nu^{syn}_{min} = \frac{\gamma_{min}^4}{4(1+z)}\nu_B$.}\\
\multicolumn{11}{l}{l: left; r: right panels of respective figure; and S: solid; D: dashed curve in figure \ref{blz-qs-flr001} }
\end{tabular}\vspace{0.3cm}\end{footnotesize}
\label{tab-fl}
\end{table*}

We find that a wide range of seed photon temperature for EC (500 -- 50k K)
capable of describing the SEDs well for both flaring and quiescent states.
Only when R/L $<$
0.1, model fails to describe SEDs (as shown in figure \ref{blz-qs-flr001}). Particularly, for high temperature 50k K, the model
fails to describe the SED for same p values (2.1) of that of the low temperature
(500 K). The modeled SED is shown by three different particle spectral indices
$p$ values 1.4, 1.8 and 2.1 (curve 1, 2 and 3 respectively in right panel of figure
\ref{blz-fl-sed}a) with lowest one ($p$=1.4) being the best description.

The estimated size L for $T_{EC}$ = 500 K is $\sim$ 3.9$\times 10^{12}$, 3.9$\times 10^{13}$ cm at SYN$^{top}_{unsc}$ = 1$\%$ (shown in left panel of figure \ref{blz-fl-sed}c, where $\lambda$ = 1.3$\times 10^{17}$cm for n$_e$ = 10$^{16}$ cm$^{-3}$, listed in table \ref{tab-fl}), and at SYN$^{top}_{unsc}$ = 10$\%$, L =6.6$\times 10^{13}$, 6.6$\times 10^{14}$ cm for R = 5L, 0.5L respectively. But if we decrease $n_e$ by factor 100, i.e.,  n$_e$ = 10$^{14}$ cm$^{-3}$ then $\lambda$ will increase with same factor, $\lambda$ = 1.3$\times 10^{19}$ cm so the estimated L is now order of $10^{14}$ cm or order of the $\gamma$-ray variability for R/L = 5 at SYN$^{top}_{unsc}$ = 1$\%$. Similarly, for SYN$^{top}_{unsc}$ = 10, 100 $\%$, with having  n$_e$ $\approx$ 2$\times$10$^{15}$, 10$^{17}$ cm$^{-3}$ respectively for R/L = 5, we can explain the observed $\gamma$-ray variability.
And for $T_{EC}$ = 50k K, the minimum electron density (i.e., for SYN$^{top}_{unsc}$ = 1$\%$) for explaining the $\gamma$-ray variability is $\sim 10^{12}$ cm$^{-3}$.
Hence to explain the 10 hours $\gamma$-ray variability of this Obs. ID. of flaring phase in this model, the electron density must be greater than 10$^{12}$ cm$^{-3}$ but the better constraint on $n_e$ will come if one can estimate observationally %
the fraction of synchrotron photons which escape from the side of the emission region (i.e., the fraction of off-axis synchrotron emission), and better constraint on $T_{EC}$.

Since broadband modeling allows us to constrain the contribution of SSC
and EC flux in total observed flux and thus, the seed photon density for SSC
(i.e., $n_e$), one can estimate the seed photon density for EC. 
The estimated range is $\sim$
(15 - 135) $n_e$ for flaring and $\sim$ (5- 60) $n_e$ for 
quiescent phase. For both phases, we are almost in single scattering limit, e.g.,
the $\langle N_{sc} \rangle$ vary from 1.10 to 1.001 for panel (a)-(c) in left
panel of figure \ref{blz-fl-sed}. The same is the case for the quiescent phase SED 
as well. 
The single scattering limit is also the reason for our
calculations not being affected from the pair-production as discussed previously.
\cite{Shah-etal2017} had modeled the above SED in one-zone model where the $\gamma_b$
is associated with peak of the two humps. On the contrary, in our continuous-jet
case mainly $\langle \gamma^2 \rangle$ (as, $\nu_p^{SSC} \approx \langle \gamma^{ran} \rangle^2 \langle \gamma^2 \rangle^2  \nu_B$) determines the $\gamma$-ray
peak. For flaring and quiescent states, $\gamma_b$ is 2500 and 1800 in \citet{Shah-etal2017} while $\sqrt{\langle \gamma^2 \rangle}$ is 1255, 706 in our continuous-jet case
(for 500K seed photon field only), comparatively lesser than one-zone model values. 
The high energy peak of modeled spectra for flaring and quiescent phase occur at $\sim$ 4$\times10^{22}$, and 4$\times10^{21}$ Hz respectively and corresponding  $\gamma_b$ is 6900 and 3881 respectively (as shown in table \ref{tab-fl}, here for both cases we fixed the same synchrotron peak frequency $\nu_p^{syn} = 1.86\times 10^{14} Hz$), hence our model is consistent with the observational fact that for high $\gamma$-ray peak we have higher synchrotron peak, which we have explicitly discussed in section \S \ref{subsec:32} (see figure \ref{blz-gen}a). In all cases (listed in table 1), the anisotropic flow is maintained as the 
  average forward velocity component 
  $\langle \gamma \rangle$ is almost more than 7 times larger to the average random velocity 
component $\langle \gamma^{ran}\rangle$ ($\approx \gamma_{min}/2$).

\subsection{Only SSC for BLLs SEDs}\label{subsec:onlySSC}
\begin{figure*} 
\centering
\begin{tabular}{lcr} \hspace{-0.5cm}
\includegraphics[scale=0.47]{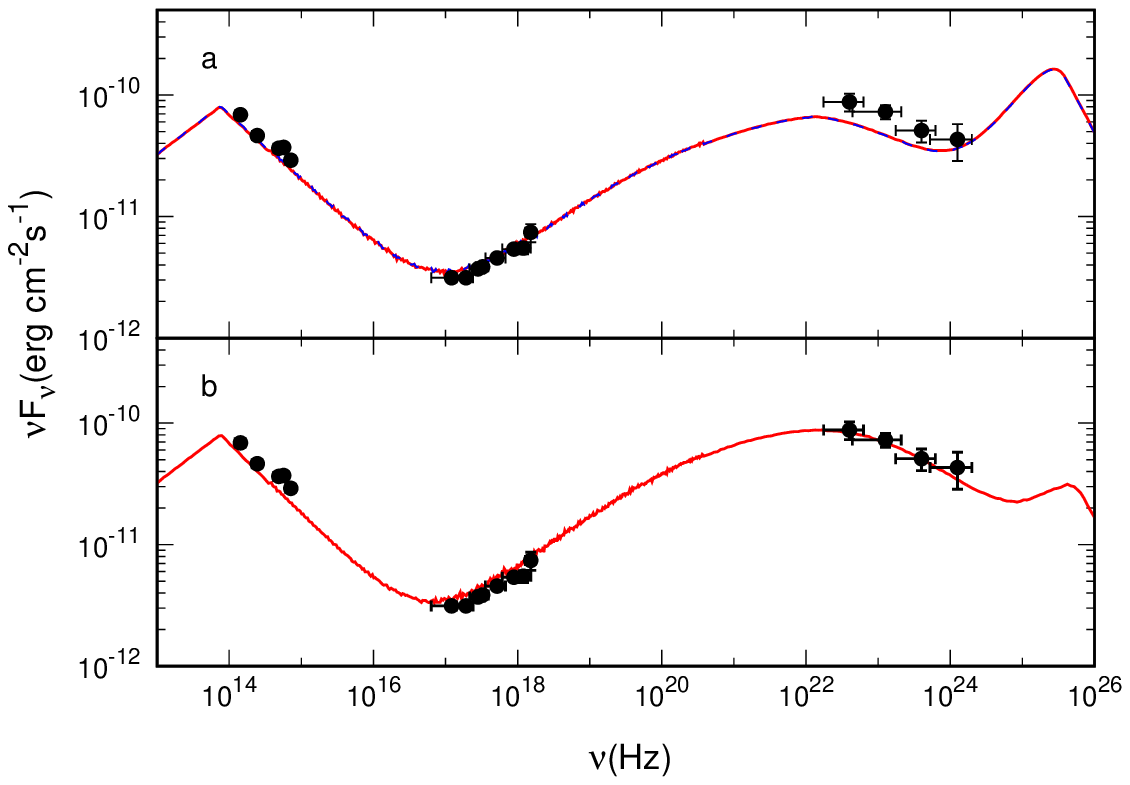}& \hspace{-.50cm}
\includegraphics[scale=0.47]{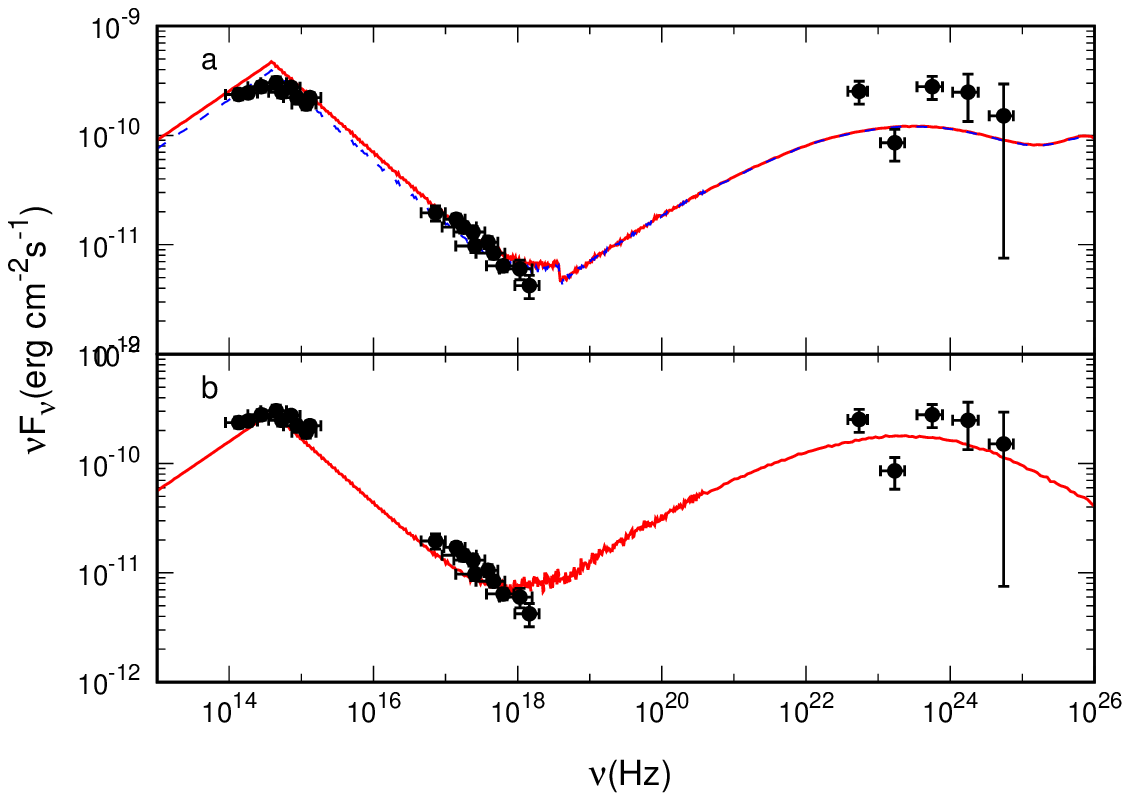}& \hspace{-.50cm}
\includegraphics[scale=0.47]{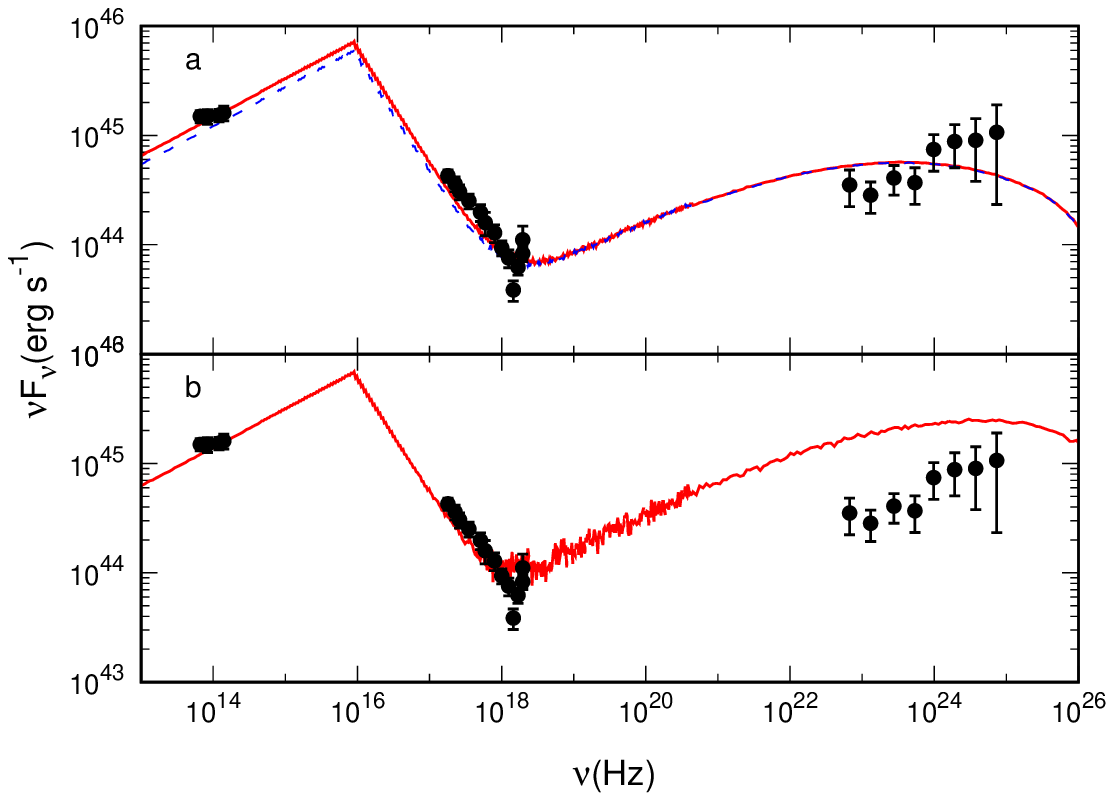}
\end{tabular} 
\caption{
  SSC explanation of high energy emission of LBL OJ 287  \citep{Kushwaha-etal2013}, IBL S5 0716+714 \citep{Chandra-etal2015},
  and HBL PKS 2155-304 \citep{Gaur-etal2017}.
  The left panel corresponds to a flaring SED of OJ 287 
the middle and right panels correspond to the IBL and HBL respectively. The curves are total emission under
synchrotron and SSC explanation (see \S \ref{subsec:onlySSC}).
}  
\label{fig:sscOj287a}
\end{figure*}

In our systematic study, 
we also explored the
parameter space to study the feasibility of SSC in reproducing high energy emission
of the LSP class of SEDs for both BLLs and FSRQs. It was based on our demonstration
in the previous section that SEDs with CD $>$ 1 
can be fitted by decreasing the R/L value. A modeled SED for BLLs OJ 287 (LBL), S5 0716+714 (IBL)  and PKS 2155-304 (HBL)
are shown in left, middle and right panel of figure \ref{fig:sscOj287a} respectively and FSRQ 3C 454.3 is %
shown in figure \ref{fig:ssc3C454} with the corresponding model parameters in the table \ref{tab:sscOj287a} and \ref{tab:ssc3C454}.

\begin{table} 
\centering
\caption{Spectral free  parameters for SED modeling (SYN+SSC) of BLLs OJ 287 (LBL), S5 0716+714 (IBL)  and PKS 2155-304 (HBL) which is shown in figure \ref{fig:sscOj287a}.  
Here, l: left; m: middle; r: right panels of figure \ref{fig:sscOj287a}; $L^o \sim 2\times10^{15}$ cm for all SEDs; rest is same as in table \ref{tab-fl}.}
\begin{footnotesize}
\begin{tabular}{|p{1.6cm}*{6}{p{0.85cm}}|}
  \hline
  & \multicolumn{2}{|c}{LBL} & \multicolumn{2}{|c}{IBL}  & \multicolumn{2}{|c|}{HBL} \\
  & \multicolumn{1}{|c}{(a)$_l$} & (b)$_l$& \multicolumn{1}{|c}{(a)$_m$} &(b)$_m$ & \multicolumn{1}{|c}{(a)$_r$} & (b)$_r$ \\ \hline
$\langle N_{sc} \rangle$   &1.61  &1.08     & 1.16  & 1.01  & 1.05 & 1.00 \\
  L (10$^{14}$cm) &5.0  &2.1     & 5.6 & 0.86 &  0.81 &  0.31 \\
  $\lambda$ (10$^{14}$cm) & 2.1 & 3.6    & 8.5 & 8.5 & 3.5  & 13.0  \\
  $B_o^{\#\#}$ (G) & 4.7e-6  & 1.6e-6   & 8.1e-7 & 8.1e-7 & 4.4e-6  & 3.2e-7  \\
  $\langle \gamma \rangle^{**}$ & 1924 & 2507    & 3828  & 3828  & 2464 & 4764 \\
  $\gamma_{b}^{**}$ & 8391 & 10973    & 25714  & 25714 & 75000 & 145000 \\
  $\gamma_{min}^{**}$ & 650 & 850   & 1150 &   1150 & 750 & 1450 \\
  $p$ & 2.1  &  2.1   & 2.1 & 2.1  & 2.3 & 2.3 \\
  $q$ & 4.1  &  4.1   & 4.1 & 4.1  & 5.1 & 5.1 \\
  R/L & 5  & 0.01 &  5 &  0.01 &  5 &  0.01  \\
  $SYN^{top}_{unsc}$($\%$) & 100   & 8   & 100 & 3 & 100  & 1\\
 \hline
\end{tabular}\vspace{0.3cm}\end{footnotesize}
\label{tab:sscOj287a}
\end{table}

In figure \ref{fig:sscOj287a}, 
the solid and dashed curves in panel (a)
represent the model flux for R/L = 5 and 0.5 respectively while the bottom panel
(b) corresponds to R/L = 0.01. 
As discussed in previous section, here also the model SED is almost
identical for
R/L = 5 and 0.5. In LBL and IBL case, CD is $\sim$ 1. We find, the best described model SED is obtained at R/L $<$ 0.01. 
Typically in both sources the $\gamma$-ray variability
time scale is order of few days \cite[e.g.,][]{Goyal-etal2018,  Chandra-etal2015}.  
The reported 2 days variability constrains the emission region size
L $\lesssim$ 2 $\times$ 10$^{15}$ cm.

For R/L =0.01,
 the model computed L is  2.1
$\times 10^{14}$ and 8.6  $\times 10^{13}$ cm for OJ 287 and S5 0716+714 respectively
(as listed in table \ref{tab:sscOj287a}), at $n_e$ = 10$^{16}$ cm$^{-3}$.
Like previously, we used the existed degeneracy between parameters $n_e$, $\lambda$, and
L (in SED modeling) to make L order of variability time scale. For this we decrease 
 $n_e$ almost by one order, i.e., $n_e$ = 10$^{15}$ cm$^{-3}$.  
The estimated $n_e$ is very large and seems unphysical. But we like to 
stress that it is a required density to generate the statistically
significant SSC spectra in single scattering limit ($\langle N_{sc} \rangle \sim $ 1).
Hence  2 days $\gamma$-ray variability of OJ 287 and S5 0716+714
can be explained with electron number density $\sim$ 10$^{15}$ cm$^{-3}$
at R/L = 0.01 and $\langle \gamma^{ran}\rangle \sim \gamma_{min}/2$.
Since we are explaining the SED for R/L $<$ 0.1
where $SYN^{top}_{unsc}$ never become 100$\%$, we also expect the emission
from the side of the jet emission region or off-axis jet emission.
As mentioned earlier, the SSC peak frequency is degenerate over $\langle
\gamma^{ran}\rangle$ and $\langle \gamma^2 \rangle$, we can model the same SED
with increasing the anisotropicity of electron motions
($\langle \gamma \rangle$).
But, we need a larger electron density, to make the emission region size
order of variability time scale, as the mean free path $\lambda$ increases with $\langle \gamma \rangle$ for given $n_e$.
In general, to explain the observed $\gamma$-ray variability of given SEDs,
the required electron density will increase with increasing the anisotropicity
of electron motions. 
For HBL PKS 2155-304, we have model spectra with CD $<$ 1. We find that for
R/L $>$ 0.01, the model SSC spectra explain the observation well with size  L
$\sim$ 8.1 $\times 10^{13}$cm at $n_e$ = 10$^{16}$ cm$^{-3}$. Similarly, 
to explain the two days %
$\gamma$-ray variability for PKS 2155-304,
the electron density should be order of 10$^{15}$ cm$^{-3}$.       

\subsection{Only SSC for FSRQ SEDs}\label{subsec:onlySSCFS}

\begin{figure} 
\centering
\begin{tabular}{lr}\hspace{-0.5cm}
  \includegraphics[width=0.46\textwidth]{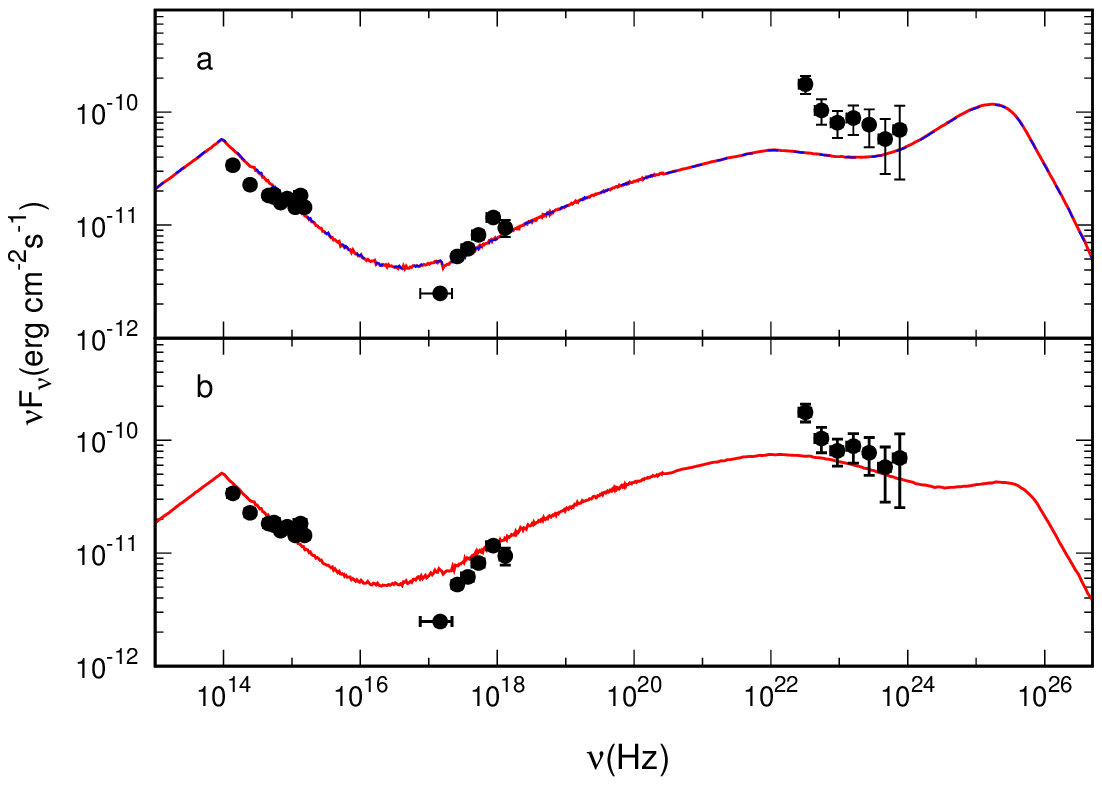}&\hspace{-.80cm}
  \includegraphics[width=0.46\textwidth]{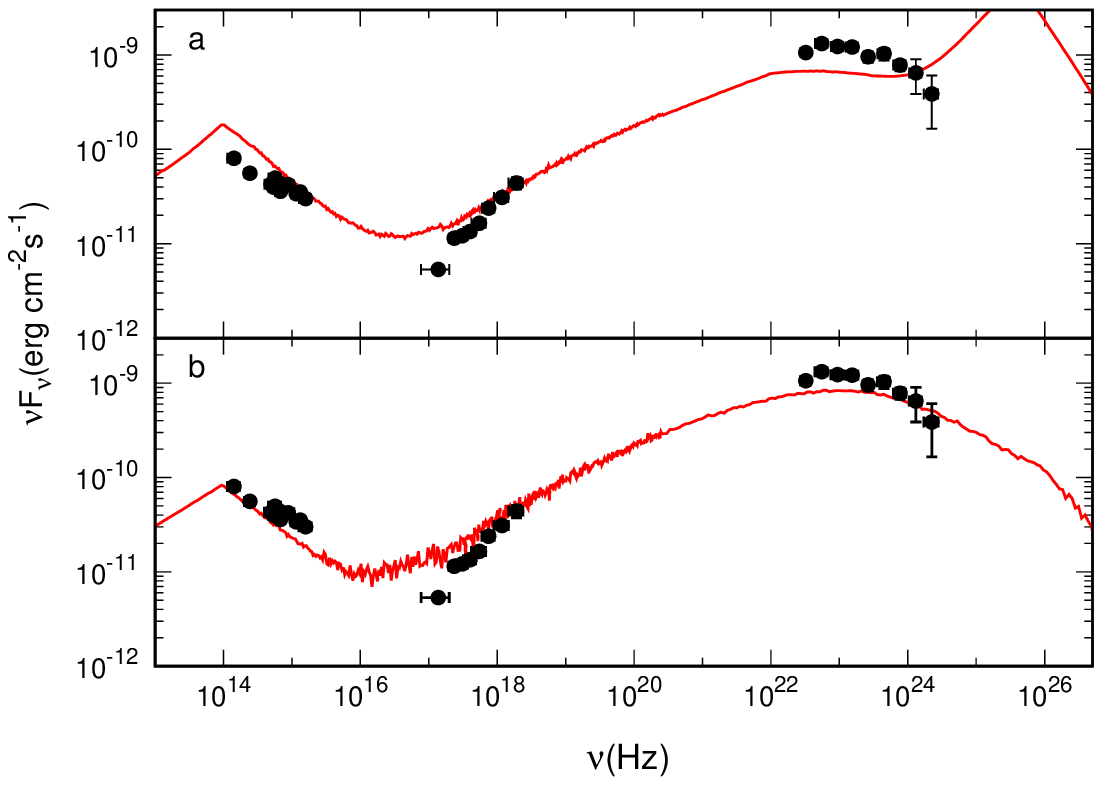}\\
\end{tabular}\vspace{-0.3cm}
\caption{Quiescent (left) and flaring (right) state SED of 3C 454.3 from
  \citep{Shah-etal2017}.
  The curves are the total emission 
  assuming synchrotron and SSC processes (see \S\ref{subsec:onlySSCFS}).
}
\label{fig:ssc3C454}
\end{figure}

In figure \ref{fig:ssc3C454},
the left and right panel show a quiescent and flaring state SED of FSRQ 3C 454.3
from \citet{Shah-etal2017} respectively.
For the quiescent state, the solid and dashed curves in panel (a) represent 
the model flux for R/L = 5 and 0.5 respectively while the
bottom panel (b)
represents model flux for R/L = 0.01.
For the flaring state, the model SSC flux is computed for R/L = 0.01
(upper panel) and 0.0001 (bottom panel) and the corresponding size  L =
8.3$\times 10^{15}$cm and = 3.3$\times 10^{15}$cm respectively at $n_e$
= 10$^{16}$ cm$^{-3}$ (as listed in table \ref{tab:ssc3C454}). In single scattering
limit, R/L $<$ 0.01 is preferred for both flaring and quiescent phase.
To explain  10 hours $\gamma$-ray variability (L $\leq$ 5$\times$10$^{14}$ cm), the electron number density
should be order of $10^{17}$ cm$^{-3}$, which is very large density even two order
greater than estimated BLLs electron density and $\sim$4--5 order greater than
EC dominated case. %
Here we like to stress that in
same analogy one can explain one second $\gamma$-ray variability with
electron density $\sim 10^{22}$ cm$^{-3}$ for this SED.
Broadly, we can produce the observed FSRQ SEDs with SYN+SSC consideration,
having a very large electron density (even comparable to the electron density of the inner disk region). However, in subsection \ref{subsec:jet_electron} we show that in this case the total electrons of the emission region is $\sim$9 orders lesser than the inner disk region. That is, the jet matters have been supplied by the inner disk region and jet matter density get increased due to a small R/L. Thus, the observed FSRQ SEDs can also be described well with  SYN+SSC consideration. 
  

\begin{table} 
\centering
\caption{Spectral free  parameters for SED modeling (SYN+SSC) of FSRQ 3C 454.3 as shown in figure \ref{fig:ssc3C454}. 
Here, l: left; r: right panels of figure \ref{fig:ssc3C454}; rest is same as in table \ref{tab-fl}.}
\begin{footnotesize}
\begin{tabular}{|p{1.6cm}*{4}{p{1.05cm}}| }
  \hline
  & (a)$_l$ & (b)$_l$&(a)$_r$ &(b)$_r$  \\ \hline
$\langle N_{sc} \rangle$   &1.77  &1.16  & 2.8  & 1.00   \\
  L (10$^{14}$cm)   &9.8  &6.3  & 83.0 &  33.0  \\
  $\lambda$ (10$^{14}$cm) &3.5  & 5.9 & 7.5 & 16.0   \\
  $B_o^{\#\#}$ (G) &7.9e-7  & 2.7e-7 & 1.7e-7 &  3.6e-8  \\
  $\langle \gamma \rangle^{**}$ &2455  &3211  & 3589  & 5289   \\
$\gamma_{b}^{**}$ & 28033  &  36658 & 40971  & 60378 \\
  $\gamma_{min}^{**}$ & 650  & 850 & 950  & 1400  \\
  $p$ & 2.1 & 2.1 & 2.1 & 2.1 \\
  $q$ & 4.1  & 4.1  & 4.1   & 4.1 \\
  R/L & 5 & 0.01  & 0.01  & 0.0001 \\
 $SYN^{top}_{unsc}$($\%$) & 100   & 10  & 25  & 0.1   \\
  \hline
  
\end{tabular}\vspace{0.3cm}\end{footnotesize}
\label{tab:ssc3C454} 
\end{table}

\subsection{Jet Power}\label{subsec:jet_pow}
The kinetic energy of the jet in leptonic case (here it is assumed that the hadrons are cold and do not participate in the radiative process, so they do not contribute to jet mechanical power significantly) 
for conical jet model is  \cite[e.g.][]{Zdziarski2014}
\begin{equation}\label{pow-jet}
  P_{jet} = \dot{M}c^2\left[\langle\gamma^R\rangle + \left(\gamma^B-1\right)\right]
\end{equation}
Here $\langle\gamma^R\rangle = \langle\gamma\rangle\langle\gamma^{ran}\rangle $, $\dot{M}$ is a mass outflow rate of jet, $\dot{M}$ = $\pi R^2 n_e m_e \beta
c$, R is the cross-section of the jet (or radius of the cylindrical emission
region), $\beta$ =  $\sqrt{1-\frac{1}{\langle\gamma\rangle^2}}$ is the
outflowing
velocity component of the electrons, $\gamma^B$ = $\left(\sqrt{1-\frac{v_{A}^2}{c^2}}\right)^{-1}$, $v_{A}$ = $B_o/\sqrt{4\pi n_e m_e}$ is the
Alfven velocity of the electrons. %
In above equation (\ref{pow-jet}), the first term represents the outflowing component and the last term %
is the magnetic field. For chosen set
of parameters, the estimated magnetic field suggest that the Alfven speed is
non-relativistic.  
Hence in presented formalism, the jet is a matter
dominated not a poynting flux \cite[e.g.,][]{Sikora-etal2005}. In addition, the contribution of hadrons in the jet power is negligible as the average Lorentz factor of electron $\langle\gamma^R\rangle$ is always greater than 5500 and hadrons are non-relativistic (cold).

 We estimated the jet power for best modeled SED of all sources. $P_{jet}$ is $\sim$2.7$\times 10^{51}$ and 7.0$\times 10^{49}$ erg/s for EC dominated (right panel of figure \ref{blz-fl-sed}c) and SSC only (right panel of figure \ref{fig:ssc3C454}b)  case of  flaring state of 3C 454.3 respectively. For BLL the jet power of
best modeled SED for OJ 287 (left panel of figure \ref{fig:sscOj287a}b), S5 0716+714 (middle panel of figure \ref{fig:sscOj287a}b)  and PKS 2155-304
(right panel of figure \ref{fig:sscOj287a}a) are $\sim$ $3.3\times 10^{52}$, $6.8\times 10^{52}$
  and $1.0\times 10^{53}$ erg/s respectively. The estimated jet power seems to be very huge.
However in presented formalism, all quantities are measured in the observers's frame, thus the radiated power $P_{rad}$ will be computed directly from the observed SED, unlike to bulk Doppler boosted model where $P_{rad}$ is measured in the comoving frame.  
%
%
%
%

Moreover the estimated jet power can be reduced by simply using the existence degeneracy between
parameters n$_e$, $\lambda$, L and R in SED modeling.
For example, we can also model the EC dominated case of above mentioned SED of 3C 454.3 for R/L=0.1 with almost
  same parameters (see figure \ref{blz-fl-sed} for R/L = 0.5) 
  and in this case the jet power becomes $\sim$ $10^{49}$ erg/s. 
In same 
way for PKS 2155-304, we can fit the SED for R/L = 0.05 with similar parameters
(see left panel of figure \ref{fig:sscOj287a} and table \ref{tab:sscOj287a}), the jet power becomes $10^{49}$ erg/s.
For considered SEDs with having jet power $P_{jet}$ $\sim 10^{49}$ erg/s the inferred R/L would be become $\lesssim$ 0.35, 0.0001, 0.0002, 0.00015 and 0.0001
for EC, SSC dominated 3C 454.3, OJ 287, S5 0716+714 and PKS 2155-304 SED
respectively. In addition the jet power can be decreased by increasing $n_e$ (which will 
lower L so R also) as P$_{jet}$ $\propto$ $n_e$R$^2$.  In general, the jet power and
$\gamma$-ray
variability time scale combinedly will give a tighter constraint on electron density and the size of the emission region. The respective constraint parameters of all considered cases are summarized in table \ref{tab:bestfitpar}.

{\bf Total averaged electrons energy in the emission region:} The total averaged electrons energy inside the emission region $W_e$ can be expressed as,
  \[W_e = \pi {\rm R^2 L} n_e \ m_e c^2 \langle \gamma^R \rangle \]
  In above paragraph, we have constrained R/L ratio by fixing $P_{jet} \sim10^{49}$ erg/s in addition to the SED modeling (listed in table \ref{tab:bestfitpar}) . We compute the total averaged electrons energies for all cases with this sets of parameter. $W_e$ = $2.2\times 10^{53}, 1.2\times 10^{54}, 8.7\times 10^{53}, 1.0\times10^{54}$ and $7.1\times10^{53}$erg
for EC, SSC dominated 3C 454.3, OJ 287, S5 0716+714 and PKS 2155-304 SED
respectively (listed in table \ref{tab:bestfitpar}).
Thus, 
within the jet-activity time scale the maximally pumped energy into the large scales structure of the jet  would not be huge, $\sim10^{54}$erg.
Hence, the estimated set of parameters is consistent with the observed large scales structure of the jet in terms of the energy budget.
Recently, \cite{Hess-2020}
have estimated the total energy in electrons $W_e = 4 \times 10^{53}$erg
for the inner, kiloparsec-scale
jet of Centaurus A  
(note, our estimated $W_e$ is for the emission region or for a core emission).
In addition, the electron cooling due to a synchrotron emission and IC emission will also reduce the total pumped energy into the large scales structure of the jet.
The electron cooling also leads to an evolution of the spectral state, 
mainly,
flaring state to the quiescent state. 
In present calculation, we did not study the evolution of the spectral state self-consistently, however the averaged electron's Lorentz factor $\langle \gamma^R \rangle$ decreases from the flaring state to quiescent (see table \ref{tab-fl} and \ref{tab:ssc3C454}).
\begin{table} 
\centering
  \caption{Constraint source parameters derived from the SED modeling and $P_{jet} \sim 10^{49}$ erg/s }
  \label{tab:bestfitpar} 
\begin{footnotesize}
  \begin{tabular}{|p{1.6cm}*{5}{p{.85cm}}| }
  \hline
Parameters  & FSRQ (EC)$^{*}$ & FSRQ (SSC)$^{*}$ & LBL$^{*}$ & IBL$^{*}$ & HBL$^{*}$  \\ \hline
  L (10$^{14}$cm)   & 5  & 5  & 20 & 20 & 20  \\
  R/L & 0.35  & 1.0e-4 & 2.0e-4 & 1.5e-4 & 2.0e-4   \\
  $n_e$  (cm$^{-3}$)$^{\dagger}$ & $10^{12}$  & $10^{17}$ & $10^{15}$ & $10^{15}$ & $10^{15}$  \\
  $n^{er}/n^{dk}$ $^{\dagger}$  & 4.0e-7  & 6.6e-9 & 4.0e-9 &  1.6e-9 & 4.0e-9\\
  $\langle \gamma \rangle^{**}$ & 321  & 5289  & 2507  & 3828 & 4764 \\
  $\langle \gamma^{ran} \rangle^{**}$ & 17.5  & 700 & 425  & 575 & 725 \\
  $p$ & 1.4  & 2.1 & 2.1 & 2.1 & 2.3 \\
  $q$ & 2.1  & 4.1 & 4.1 & 4.1 & 5.1 \\
  $B_o$ (G) & 1.0e-2 & 3.6e-8 & 1.6e-6 & 8.1e-7 & 3.2e-7 \\
  $P_{jet}$ (erg/s) & 1.3e49 & 7.1e49 & 1.3e49 & 1.5e49 & 1.0e49 \\
  $W_e$ (erg) & 2.2e53  & 1.2e54  & 8.7e53   & 1.0e54 & 7.1e53 \\
\hline
\multicolumn{6}{r}{\parbox[t]{8.1cm}{ $^{*}$ FSRQ, LBL, IBL and HBL  stand for sources  3C 454.3  \citep{Shah-etal2017}, OJ 287  \citep{Kushwaha-etal2013},  S5 0716+714 \citep{Chandra-etal2015},  and PKS 2155-304 \citep{Gaur-etal2017} respectively. EC and SSC are for EC and SSC dominated cases for 3C 454.3 respectively.}  }  \\
\multicolumn{6}{r}{\parbox[t]{8.1cm}{$^{\dagger}$ In some cases $n_e$ becomes comparable to the inner region of the disk, however the total electrons of the emission region is  always 7 orders (i.e., $n^{er}/n^{dk} \lesssim 10^{-7} $) lesser than the inner disk region. (see text for details)  }}  \\
\multicolumn{6}{l}{$^{**}$ in units of $m_e c^2$}\\
%
  \end{tabular}\vspace{0.3cm}\end{footnotesize}
\end{table}

\subsection{Estimated $n_e$ of the jet emission region and matter supplied to the jet from the base of the jet}\label{subsec:jet_electron}

With constraint of optically thin medium for IC, of the $\gamma$-ray variability time scale, and of a statistical significant modeled IC spectra, the estimated electron density of the jet emission region is appeared as a large quantity especially for BL Lac objects or only SSC cases.
  Here, we examine the physicality of large electron density in emission region by estimating 
the total injected matters into the jet emission region from the inner region of the accretion disk (or from the base of the jet) . 
For this, we compute the total number of electrons inside the emission region of the jet ($n^{er}$) and inside the inner region of the accretion disk ($n^{dk}$). The volume of the inner region of the disk is $\pi r_{in}^2 h$, here $r_{in}$ is the outer radius of the inner region, and $h$ is the disk scale height. 
Typically, $r_{in}$ is the order of $10 - 30 R_g$ for the base of the jet \cite[e.g.][]{Mckinney2006, Romero-etal2017}. 
The ratio of the total number of electrons inside the emission region to inside the inner region of the disk is expressed as,
\begin{equation}
 \frac{n^{er}}{n^{dk}} = \frac{{\rm R^2L}n_e}{r_{in}^2 h\ n_e^{dk}} \approx \frac{10^3{\rm R^2L}n_e}{r_{in}^3 n_e^{dk}} 
\end{equation}
Above, for approximate expression we use the relation of thin accretion disk (\cite[][]{Shakura-Sunyaev1973}) $r/h \sim 10^{3}$ at $r \sim r_{in}$, and $n_e^{dk}$ is the electron density of the disk at $r_{in}$.
Mostly, the mass of the central black hole in blazars is uncertain.
For a rough estimation of the ratio $\frac{n^{er}}{n^{dk}}$ in case of BL Lac objects, we consider the mass of the black hole
$M_{BH} = 10^8 M_\odot$, which gives
$n_e^{dk} \sim 10^{18}$ cm$^{-3}$ at $r=30R_g$ for a mass accretion rate $\dot{M} = 10^{-2} \dot{M}_{Edd}$ in a thin accretion disk, here $\dot{M}_{Edd}$ is the
Eddington mass accretion rate.
From the calculations (see subsection \ref{subsec:onlySSC}),  we have L $\sim 2 \times 10^{15}$ cm $\approx 100 R_g$,  $n_e = 10^{15}$ cm$^{-3}$, and R/L $\lesssim 10^{-2}$ for the BL Lac objects.
Thus, $\frac{n^{er}}{n^{dk}} \approx 10^{-4}$ and $10^{-6}$ for R/L = $10^{-2}$ and $10^{-3}$ respectively (here, $r_{in}$=L).
Crudely, the total number of electrons of emission region of BL Lac
($M_{BH} =10^8 M_\odot$) is always 6 orders lesser than the total electrons
of the inner disk region provided R/L $\lesssim 10^{-3}$.

Masses of the central black hole of 3C 454.3, S5 0716+714 and PKS 2155-304 are uncertain and have a wide ranges, which are (0.3 - 2)$\times 10^9 M_\odot$ \cite[][]{Bonnoli-etal2011}, (0.3 - 4)$\times 10^9 M_\odot$ \cite[][]{Kaur-etal2018} and (0.3 - 3)$\times 10^9 M_\odot$ \cite[][]{Gupta2014} respectively. The primary black hole mass of OJ 287 (a binary supermassive black hole system) is $\sim$1.8 $\times 10^{10} M_\odot$ \cite[][and references therein]{Ramirez-etal2020}.
Unlike the rough estimation of the ratio $n^{er}/n^{dk}$ (in above paragraph), here we estimate this ratio for constraint parameter sets obtained by SED modeling and $P_{jet}$. The constraint R/L is 0.35, 0.0001, 0.0002, 0.00015 and 0.0002 and $n_e$ is $10^{12}, 10^{17}, 10^{15}, 10^{15}$ and $10^{15}$ cm$^{-3}$ for EC and SSC dominated 3C 454.3, OJ 287,  S5 0716+714 and PKS 2155-304 respectively for $P_{jet} \sim 10^{49}$ erg/s (see section \S\ref{subsec:jet_pow} or table \ref{tab:bestfitpar}).
To compute $n^{er}/n^{dk}$, we take an average value of the black hole mass for all sources $M_{BH} = 10^9M_\odot$, and $r_{in} = 30R_g$, with this the electron density of the inner region of the disk  is $n_e^{dk} \sim 3\times 10^{17}$ cm$^{-3}$.
The emission region size 
L is $\sim 3R_g$ and $\sim 10R_g$ for considered FSRQ and BL Lac objects respectively.
The estimated ratio of electron in emission region to inner region of
the disk is $\frac{n^{er}}{n^{dk}}$ = $\frac{0.12 n_e}{n_e^{dk}}$, $\frac{2\times 10^{-8} n_e}{n_e^{dk}}$, $\frac{1.2\times10^{-6} n_e}{n_e^{dk}}$, $\frac{6.6\times10^{-7} n_e}{n_e^{dk}}$, $\frac{1.2\times10^{-6} n_e}{n_e^{dk}}$
for EC and SSC dominated 3C 454.3, OJ 287,  S5 0716+714 and PKS 2155-304 respectively (see, table \ref{tab:bestfitpar}).
In summary, though the estimated $n_e$ seems huge but due to a small R/L
the total electrons present
in emission region size is many orders  smaller than the inner region of
the disk. That is, the matter of jet emission region has been injected by the inner accretion disk around the central black hole, and thus the estimated electron density of jet's emission region is consistent in terms of the  disk-jet connection.
%

\section{Summary and Conclusions}\label{sec:sumCon}
Based on the observed jet morphology and a wide range of rarely repeating spectral,
temporal, polarization behaviors, we explored blazars broadband SEDs in a locally
anisotropic, relativistic flow within a steady conical jet assuming leptonic
emission scenario.
The highly relativistic velocity along the jet axis dominates to its
  random perpendicular component. 
The anisotropically moving electron (along the jet axis) radiates synchrotron emission in forward direction (thus beamed in the forward
direction) in presence of helical wiggler plus axial magnetic field where electron follows the helical path without changing its axial velocity component and wiggler field directed random component.
Assuming a broken-powerlaw energy distribution of the electrons, we calculate
the synchrotron and IC emission from a cylindrical emission region of radius R and
length L. For the latter, we employ Monte Carlo method. We emphasized that in single
scattering limit the photon-photon scattering for pair production does not dominate over
IC process. The model SED has a best description of observation when it is generated in
single scattering limit. We performed modeling of SEDs of different class of blazars 
(BLLs: OJ 287 (LBL), S5 0716+714 (IBL), PKS 2155-304 (HBL); FSRQ: 3C 454.3))
in their different flux states and also explore the effects of geometry on the
parameters and observed emission. The observed short $\gamma$-ray variability has
been explained in terms of electron density, since the free parameters $n_e$, $\lambda$,
L are degenerate in the IC process. For example, ten hours variability of flaring
state of 3C 454.3 is explained with $n_e$ $\gtrsim$ 10$^{12}$ cm$^{-3}$ in EC dominated
case (also we have estimated the seed photon density for EC, which is $\sim$(15-135) $n_e$).
Two days variability of OJ 287 is described with high $n_e$ $\gtrsim$ $10^{15}$
cm$^{-3}$ for $\langle \gamma^{ran} \rangle$ = $\gamma_{min}/2$.
However, this apparently large $n_e$ is due to the requirement that
SSC reproduces the observed spectra in single scattering limit.
Moreover, we have shown that the total electrons present
in emission region size is always 7 orders smaller than the inner region of
the disk, i.e., the matter of jet emission region has been injected by the inner accretion disk around the central black hole.

We also explored the effect of dimension on the observed broadband spectrum
and found that high energetic emission of LSP blazar can be explained through
SSC only by decreasing the radius to length ratio (R/L) of cylindrical emission
region (R/L $< 0.001$). However, this also results in a significant fraction of
unscattered photons leaving from side rather than the front face of the cylindrical
emission
region. For example, the flaring phase of 3C 454.3 has almost an order of magnitude
more emission at $\gamma$-ray compared to the synchrotron part and we can model this
SED considering SSC only. The estimated R/L is $\lesssim$ 0.0001 and the associated
10 hours variability is explained with $n_e$ $\gtrsim$ 10$^{17}$ cm$^{-3}$. We find
that for all considered SEDs, the jet is a matter dominated. The estimated jet power
is consistent with observed radiative power and constrain tightly the emission region
size and electron density along with $\gamma$-ray variability.
Conclusively, in a
steady  conical jet assuming leptonic emission scenario and anisotropic
flow in helical periodic  magnetic field plus axial field, one can modeled the blazar SED successfully; the
radiation is beamed, but not by bulk Doppler boosting, and is unaffected from
the pair-production. 

\section*{Acknowledgements}
NK is supported by University Grant Commission (UGC), New Delhi, India through Dr. D.S. Kothari Post-Doctoral Fellowship (201718-PH/17-18/0013). 
NK thanks Banibrata Mukhopadhyay of IISc for discussion.
NK likes to acknowledge his IUCAA visit (April 2019) and thanks Dipankar Bhattacharya and Ranjeev Misra for critical comments and suggestions over the manuscript.

\subsection*{Data availability}
The data is taken from following published work --

\noindent[Figure: \ref{blz-demo-I}] \cite{Kushwaha-etal2018a}

\noindent[Figures: \ref{blz-demo-II}a; \ref{blz-fl-sed}; \ref{blz-qs-sed}; \ref{blz-qs-flr001}; \ref{fig:ssc3C454} ] \cite{Shah-etal2017}           

\noindent[Figures: \ref{blz-demo-II}b \ref{fig:sscOj287a}a] \cite{Kushwaha-etal2013}

\noindent[Figure: \ref{fig:sscOj287a}b] \cite{Chandra-etal2015}           

\noindent[Figure: \ref{fig:sscOj287a}c] \cite{Gaur-etal2017}

\def\aap{A\&A}%
\def\aapr{A\&A~Rev.}%
\def\aaps{A\&AS}%
\def\aj{AJ}%
\def\actaa{Acta Astron.}%
\def\araa{ARA\&A}%
\def\apj{ApJ}%
\def\apjl{ApJ}%
\def\apjs{ApJS}%
\def\apspr{Astrophys.~Space~Phys.~Res.}%
\def\ao{Appl.~Opt.}%
\def\aplett{Astrophys.~Lett.}%
\def\apss{Ap\&SS}%
\def\azh{AZh}%
\def\bain{Bull.~Astron.~Inst.~Netherlands}%
\def\baas{BAAS}%
\def\bac{Bull. astr. Inst. Czechosl.}%
\def\caa{Chinese Astron. Astrophys.}%
\def\cjaa{Chinese J. Astron. Astrophys.}%
\def\fcp{Fund.~Cosmic~Phys.}%
\def\gafd{Geophys.\ Astrophys.\ Fluid Dyn.}
\def\gca{Geochim.~Cosmochim.~Acta}%
\def\grl{Geophys.~Res.~Lett.}%
\def\iaucirc{IAU~Circ.}%
\def\icarus{Icarus}%
\def\jcap{J. Cosmology Astropart. Phys.}%
\def\jcp{J.~Chem.~Phys.}%
\def\jfm{JFM}
\def\jgr{J.~Geophys.~Res.}%
\def\jqsrt{J.~Quant.~Spec.~Radiat.~Transf.}%
\def\jrasc{JRASC}%
\def\mnras{MNRAS}%
\def\memras{MmRAS}%
\def\memsai{Mem.~Soc.~Astron.~Italiana}%
\def\na{New A}%
\def\nar{New A Rev.}%
\def\nat{Nature}%
\def\natas{Nature Astronomy}%
\def\nphysa{Nucl.~Phys.~A}%
\def\pasa{PASA}%
\def\pasj{PASJ}%
\def\pasp{PASP}%
\def\physrep{Phys.~Rep.}%
\def\physscr{Phys.~Scr}%
\def\planss{Planet.~Space~Sci.}%
\def\pra{Phys.~Rev.~A}%
\def\prb{Phys.~Rev.~B}%
\def\prc{Phys.~Rev.~C}%
\def\prd{Phys.~Rev.~D}%
\def\pre{Phys.~Rev.~E}%
\def\prl{Phys.~Rev.~Lett.}%
\def\procspie{Proc.~SPIE}%
\def\qjras{QJRAS}%
\def\rmxaa{Rev. Mexicana Astron. Astrofis.}%
\def\sgg{Stud.\ Geoph.\ et\ Geod.}
\def\skytel{S\&T}%
\def\solphys{Sol.~Phys.}%
\def\sovast{Soviet~Ast.}%
\def\ssr{Space~Sci.~Rev.}%
\def\zap{ZAp}%
\def\memsai{Memorie della Societa Astronomica Italiana}

\bibliographystyle{aa}
\bibliography{pap5}

\appendix

  \section{Synchrotron emission by anisotropicaly moving electron in wiggler plus axial magnetic field}
 
We are revisiting the single relativistic electron trajectory in the periodic helical (wiggler) magnetic field and axial filed \cite[studied by][]{Friedland1980, Freund-Drobot1982} and discussed the synchrotron emission by electron with application to our scenario.
  The combined magnetic field, axial (of amplitude $B_o$) plus wiggler (of amplitude $B_w$) is defined as
  \[ \text{\bf B} = B_o  \text{\bf k} + B_w (\text{\bf i}\cos{(k_w z)}+\text{\bf j}\sin{(k_w z)} )\]
here, $2\pi/k_w$ is the wiggler period; (\textbf{i,j,k}) are the cartesian axis where \textbf{k} is along the jet-axis. The velocity components of electron are $\rm v_1, v_2, v_3$ where v$_3$ is along the \textbf{k}-axis and $v_1$ is along the wiggler field direction while v$_2$ is perpendicular to $v_1$ in (\textbf{i},\textbf{j})-plane. The equation of motion is written as,
\[ \dot{\rm v_1} = {\rm v_2}(k_w {\rm v_3} - \Omega_o); \quad \dot{\rm v_2} = -\Omega_w {\rm v_3 - v_1} (k_w {\rm v_3}-\Omega_o); \quad \dot{\rm v_3}=\Omega_w \rm{v_2} \]
where $\Omega_o = \frac{eB_o}{\gamma m_e c}$, $\Omega_w = \frac{eB_w}{\gamma m_e c}$, $\gamma$ = $\left(1-\frac{\rm v^2}{c^2}\right)^{-1/2}$, $\rm v^2=v_1^2+v_2^2+v_3^2$, and $\rm\dot{v}_i$ is the time derivative of velocity $\rm v_i$. The equation of motion obeys two constant of the motion, i) the total energy is conserved, i.e., $\rm v^2$ = constant; ii) ${\rm u = v_1} - k_w \frac{({\rm v_3}-\frac{\Omega_o}{k_w})^2}{(2\Omega_o)}$ = constant. By using the constant of motion the equation of motion will  reduce to
\[ \dot{\rm v_3}^2 = \phi(\rm v_3,v,u)\]
where $\phi(\rm v_3,v,u)$ is a quartic equation for variable v$_3$. The real roots of $\phi$ correspond to the solution v$_3$ = constant, v$_1$ = $\frac{-\Omega_w {\rm v_3}}{k_w {\rm v_3} - \Omega_o}$, and v$_2$ = 0 (termed as constant axial velocity solution). The electron follows the wiggler trajectory and radiates synchrotron emission in forward direction.

In above description a non-relativistic energy equation has been considered, as a result in constant axial velocity solution the both velocity components v$_1$ and v$_3$ cannot be relativistic together, i.e., for relativistic v$_3$ the other velocity component v$_1$ would be non-relativistic and vice-varsa. But when one considers a relativistic energy equation (i.e., $\rm {v^2 = v_1^2 + v_3^2 + v_2^2} \frac{1}{\gamma_1^2 \gamma_3^2}$ where $\gamma_1$, $\gamma_3$ are the Lorentz factor for velocity v$_1$ and v$_3$ respectively) then both v$_3$ and v$_1$ can be relativistic, which is our considered case. Thus, the both velocity components of electron can be relativistic in this filed configuration \cite[e.g.,][]{Pellegrini-etal2016}.



\end{document}